\documentclass[reprint, amsmath, amssymb, aps, nofootinbib]{revtex4-2}
\usepackage{graphicx}
\usepackage{dcolumn}
\usepackage{bm}
\usepackage{hyperref}
\hypersetup{
    colorlinks=true,
    linkcolor=blue,
    filecolor=blue,      
    urlcolor=blue,
    citecolor=blue
    }
\usepackage{amsmath,amssymb,amsfonts,bm}
\usepackage{ulem}
\usepackage[dvipsnames]{xcolor}

\begin{document}


\title{Higher order flow coefficients - A Messenger of QCD medium formed in heavy-ion collisions at the Large Hadron Collider}

\author{Suraj Prasad}\email[]{Suraj.Prasad@cern.ch}
\author{Aswathy Menon Kavumpadikkal Radhakrishnan}\email[]{Aswathy.Menon@cern.ch}
\author{Raghunath Sahoo}\email[Corresponding author: ]{Raghunath.Sahoo@cern.ch}
\affiliation{Department of Physics, Indian Institute of Technology Indore, Simrol, Indore 453552, India}
\author{Neelkamal Mallick}\email[]{Neelkamal.Mallick@cern.ch}
\affiliation{University of Jyv\"askyl\"a, Department of Physics, P.O. Box 35, FI-40014, Jyv\"askyl\"a, Finland}

\date{\today}

\begin{abstract}
Anisotropic flow and fluctuations are sensitive observables of the initial state effects in heavy ion collisions and are characterized by the medium properties and final state interactions. Using event-shape observables, one can constrain the probability distributions of anisotropic flow coefficients, thus reducing the linear and nonlinear contributions in the measured higher-order harmonics. In this paper, we use transverse spherocity as an event shape observable to study the flow coefficients and elliptic flow fluctuations. Transverse spherocity is found to have a strong correlation with elliptic flow and its fluctuations. We exploit this feature of transverse spherocity to remove the contribution to elliptic flow from higher-order harmonics. The study is performed in Pb--Pb collisions at $\sqrt{s_{\rm NN}}=5.02$ TeV using a multi-phase transport model. The multi-particle Q-cumulant method estimates the anisotropic flow coefficients, which reduces the non-flow contributions. We observe a stronger system response to the flow coefficients for the events with smaller values of elliptic flow.

\end{abstract}

\maketitle

\section{Introduction}
\label{sec:intro}

Heavy-ion collisions at the Large Hadron Collider, CERN, Switzerland, and the Relativistic Heavy-Ion Collider at BNL, USA, create laboratory conditions for the formation of a hot and dense plasma of quarks and gluons, which is believed to have existed in the microsecond-old Universe shortly after the Big Bang~\cite{Bass:1998vz}. Under the extreme conditions of high temperature and energy densities achieved in such collisions, it is feasible to test and study the theory of strong interactions via quantum chromodynamics (QCD). Nonetheless, the quark-gluon plasma (QGP) is very short-lived; thus, a direct observation of such a medium, if formed, is elusive. Consequently, several signatures of QGP can be measured in the experiments, which include strangeness enhancement~\cite{Rafelski:1982pu}, jet quenching~\cite{Appel:1985dq}, quarkonia suppression~\cite{Matsui:1986dk}, collective flow~\cite{Herrmann:1999wu}, etc. While the presence of collective flow hints at the formation of a hydrodynamically expanding medium of deconfined partons, the measurement of anisotropic flow coefficients of the final state hadrons underlines several crucial properties of the medium formed. The anisotropic flow can be quantified in terms of the coefficients of the Fourier expansion of the final state azimuthal distribution of the emitted particles, as follows~\cite{Voloshin:1994mz}:
\begin{equation}
    \frac{dN}{d\phi}\propto\;1+2\sum_{n=1}^{\infty}v_{n}\cos[n(\phi-\psi_{n})]
    \label{eq:Fourier}
\end{equation}
Here, $N$ is the number of particles, $\phi$ is the azimuthal angle of emission, and $n$ denotes the order of the harmonics. $\psi_{n}$ is the $n^{\rm th}$ order symmetry plane angle. $v_{n}$ denotes the $n$th order anisotropic flow coefficient, where $v_1$, $v_2$, and $v_3$ denote the directed flow, elliptic flow, and triangular flow, respectively~\cite{Heinz:2013th}. The anisotropic flow coefficients depend on several factors, which include the initial spatial anisotropy, initial density fluctuations, and bulk and transport properties of the medium. Because the nuclear overlap is approximately elliptical, the average geometry dominates $v_2$.  On the other hand, $v_n$ for $n > 2$ is dominated by geometrical fluctuations or, more precisely, the position fluctuations of the individual nucleons, which in turn generate geometrical fluctuations.

The initial spatial anisotropy of the collision overlap region can be quantified in terms of initial eccentricities ($\epsilon_{n}$), where, for $n=2$ and 3, the contributions to  $v_{n}$ from corresponding $\epsilon_n$ are linear\footnote{See Eq.~\eqref{eq:eccentricity} for explicit definition of $\epsilon_{n}$.}, \textit{i.e.}, $v_{2}\propto\epsilon_2$ and $v_{3}\propto\epsilon_3$. 
However, harmonic flow coefficients of higher order ($n>3$) arise not only from the initial eccentricities of the same order but also from the non-linear mixing of lower-order harmonics.  
In fact, $v_{4}\{2\}$ has a linear contribution from $\epsilon_4$, and a nonlinear contribution from $\epsilon_2$ or $v_2$, i.e.~\cite{Teaney:2013dta, Gardim:2011xv},
\begin{eqnarray}
    v_4 e^{i4\Phi_4}=a_0 \epsilon_4e^{i4\psi_4}+a_1(\epsilon_2e^{i2\psi_2})^2+\dots \\ \nonumber
    =c_0 e^{i4\psi_4}+c_1(v_2e^{i2\Phi_2})^2+\dots,\;\;
\end{eqnarray}
where $\Phi_n$ is the phase, known as the event plane. $c_0=a_0\epsilon_4$ is the linear component of $v_4$. It is important to note that the coefficients, $a_0$, $a_1$ and $c_1$ vary weakly with change in centrality. These nonlinear contributions of $v_{2}$ in the measured value of $v_{4}$ can lead to strong centrality dependence of the correlation between $\psi_{2}$ and $\psi_{4}$~\cite{Teaney:2013dta}. In addition, $v_5$ has a linear contribution from $\epsilon_5$ and a nonlinear contribution from both $v_2$ and $v_3$~\cite{Teaney:2013dta, Gardim:2011xv}:
\begin{eqnarray}
    v_5 e^{i5\Phi_5}=a_0 \epsilon_5e^{i5\psi_5}+a_1\epsilon_2e^{i2\psi_2}\epsilon_3e^{i3\psi_3}+\dots \\ \nonumber
    =c_0 e^{i5\psi_5}+c_1v_2v_3e^{i(2\Phi_2+3\Phi_3)}+\dots,
    \label{eq:v5vsv2v3}
\end{eqnarray}
which leads to the observed event plane correlations involving $\psi_2$, $\psi_3$, and $\psi_5$ in the experiments~\cite{ATLAS:2014ndd} using the event plane method, which is absent for correlations estimated using the Gaussian estimator method~\cite{ALICE:2024fus}. These findings are crucial because, in hydrodynamical models, it is observed that the anisotropic flow coefficients are affected significantly by a change in specific shear viscosity ($\eta_{s}/s$) of the medium and the impact of $\eta_{s}/s$ on $v_{n}$ increases with increasing $n$~\cite{Chaudhuri:2011pa, Schenke:2011bn}. Thus, higher-order anisotropic flow coefficients have proven to be more effective probes to estimate the transport coefficients of the medium formed in relativistic collisions. Consequently, it is important to decouple the measured signal of the $v_2$ and $v_3$ from the higher-order harmonics to have a better estimate of the properties of the medium formed, and here comes into play the role of event shape estimators~\cite{ATLAS:2015qwl}.

The values of $\epsilon_n$ and consequently $v_{\rm n}$ can fluctuate event-by-event due to the fluctuations of the participating nuclear positions at the initial state~\cite{ALICE:2022wpn}. The distributions of $v_{n}$ are very broad, even in fine centrality bins, where the values of $v_{2}$ and $v_{3}$ are found to fluctuate from zero to several times their mean values~\cite{ATLAS:2013xzf}. These fluctuations in the measurements of $v_{2}$ and $v_{3}$ can, in turn, affect the measurements of higher-order flow coefficients. Therefore, by cleanly selecting the events with different $v_{2}$ and $v_{3}$ values, one can constrain their contributions in the higher-order flow harmonics. This makes the study of event shapes important in heavy-ion collisions, where one can select events based on their topology, which provides an independent method to separate the nonlinear contributions of $v_{2}$ and $v_{3}$ in higher-order flow harmonics.

Traditionally, the event shapes are determined using the $n$th order reduced flow vectors ($q_{n}$), which is a multiplicity normalised value of the flow vector ($Q_{n}$), defined as follows~\cite{STAR:2002hbo, Schukraft:2012ah}:
\begin{equation}
    q_n=\frac{|Q_{n}|}{\sqrt{M}},
    \label{eq:qn}
\end{equation}
where,
\begin{equation}
      Q_{n}=\sum_{j=1}^{M}e^{in\phi_{j}}
\label{eq:Qn}
\end{equation}
Here, $M$ is the total number of charged particles in a given rapidity region, $\phi_{j}$ is the azimuthal angle of the $j$th charged particle, and $i=\sqrt{-1}$ is the imaginary unit. $q_n$ vectors are proportional to the anisotropic flow coefficients for a large particle multiplicity limit. However, towards the peripheral collisions, where the particle multiplicity is small, $q_n$ are not good observables to separate the elliptic or triangular events based on the final state azimuthal distribution of charged particles in the transverse plane~\cite{ATLAS:2015qwl}. Further, one should be careful when dealing with low multiplicity events, where a deviation of linear $\epsilon_2$ scaling to $v_2$ is expected due to a cubic response term~\cite{Noronha-Hostler:2015dbi}. In contrast, transverse spherocity ($S_0$), defined in Sec.~\ref{sec:sphero}, is an event shape observable traditionally used in small systems like proton-proton ($pp$) collisions to select the soft events dominated by non-pQCD interactions from the pQCD-dominated jetty events~\cite{ALICE:2019dfi}. In addition, recent simulation studies show that transverse spherocity can also be used as an event shape observable in heavy-ion collisions~\cite{Mallick:2021hcs, Prasad:2021bdq, Prasad:2022zbr, Mallick:2020ium, Tripathy:2025npe}. It has previously been observed that transverse spherocity has a significant correlation with both elliptic and triangular flow in addition to the $q_2$ and $q_3$ vectors~\cite{Prasad:2022zbr}. Interestingly, transverse spherocity-based event selections are observed to affect the number of constituent quark (NCQ) scalings of the elliptic flow coefficients of identified hadrons~\cite{Mallick:2021hcs}. Additionally, one observes that the strength of the symmetry plane correlations changes with different classes of transverse spherocity~\cite{Tripathy:2025npe}. Furthermore, the events with large values of $S_0$ show a large mean transverse radial flow velocity and a smaller kinetic freezeout temperature in Pb--Pb collisions compared to events with a smaller $S_0$ value~\cite{Prasad:2021bdq}. The studies mentioned above signify both the applicability and importance of event shape selection based on transverse spherocity in heavy-ion collisions.

We take this opportunity to exploit transverse spherocity to study the higher-order anisotropic flow coefficients and elliptic flow fluctuations. The study is performed in Pb--Pb collisions at $\sqrt{s_{\rm NN}}=5.02$ TeV using a multi-phase transport (AMPT) model. Studying elliptic flow fluctuations with transverse spherocity would help in selecting events with smaller elliptic fluctuations and thereby constrain the value of $v_{2}$. Consequently, a more meaningful study of higher-order flow coefficients can be performed.

The paper is organised as follows. We start with a brief introduction in Sec.~\ref{sec:intro}, followed by discussions of event generation and methodology in Sec.~\ref{sec:eventgen}. In Sec.~\ref{sec:results}, the results and corresponding discussions are presented. Finally, in Sec.~\ref{sec:summary}, a brief summary and outlook of the study are given.

\section{Event generation and methodology}
\label{sec:eventgen}

In this section, we begin with a brief discussion on event generation using a multi-phase transport model, followed by the definition and calculation of transverse spherocity. The multi-particle cumulant method used to estimate the flow coefficients $v_{\rm n}$ is described in brief thereafter.

\subsection{A multi-phase transport model}

The AMPT model is a Monte Carlo transport model that provides a kinetic description of the various stages involved in heavy-ion collisions. Collision initialization, parton transport, hadronization, and hadronic transport are the four phases incorporated in the AMPT model~\cite{Zhang:1999bd, Lin:2004en}. The initial parton distributions are generated using the heavy-ion jet interaction generator (HIJING)~\cite{Wang:1991hta}. The interactions among the partons are simulated using Zhang's parton cascade (ZPC)~\cite{Zhang:1997ej}. Conversion of partons to hadrons is done via the Lund string fragmentation model (in default AMPT)~\cite{Andersson:1983ia} or through the quark coalescence mechanism (in the string-melting version of AMPT)~\cite{Greco:2003xt}. In the final stage, the formed hadrons undergo a final evolution in a relativistic transport (ART) model framework via baryon-baryon, meson-baryon, and meson-meson interactions~\cite{Li:1995pra,Li:2001xh}.

As this work focuses on studying the flow coefficients and flow fluctuations, we use the string-melting (SM) mode (version 2.26t9b) of AMPT wherein the quark coalescence mechanism of hadronization is known to describe well the particle $p_{\rm T}$-spectra and flow at intermediate $p_{\rm T}$~\cite{Greco:2003mm,Fries:2003vb,Fries:2003kq, Mallick:2023vgi, Mallick:2022alr}. The AMPT settings used in this study are the same as reported in Ref.~\cite{Tripathy:2018bib}.

\subsection{Transverse spherocity}
\label{sec:sphero}
Well-known for its ability to categorize events based on their azimuthal topology, the variable, transverse spherocity ($S_{0}$) is defined for an event as~\cite{Banfi:2010xy, Farhi:1977sg, Cuautle:2014yda, ALICE:2019dfi, Prasad:2025yfj}:
\begin{equation}
    S_{0}=\frac{\pi^2}{4}\min_{\hat{n}}\Bigg(\frac{\sum_{i=1}^{N_{\rm had}}|\bf{p_{\rm T}}_{i} \times \hat{n}|}{\sum_{i=1}^{N_{\rm had}}|\bf{p_{\rm T}}_{i}|}\Bigg)^{2}
    \label{eq:spherodefn}
\end{equation}

where $\bf{p_{\rm T}}_{i}$ is the transverse momentum vector of $i^{\rm th}$ charged hadron in the event, and the unit vector $\hat{n}$ is to be chosen such that it minimizes the term in the bracket. Here, the minimization is carried out over all possible angles in the azimuth in the transverse plane of the event, ranging from 0 to $2\pi$. Events with a minimum of 5 charged particles are chosen for estimating $S_{0}$, and the charged particles in the range $|\eta|< 0.8$ with $p_{\rm T} > 0.15$ GeV/$c$ alone enter the calculation of $S_{0}$. The normalization constant $\pi^{2}$/4 ensures that $S_{0}$ values range between 0 and 1. While the lower spherocity limit $S_{0} = 0$ characterizes an event with pencil-like emission of particles called the jetty event, $S_{0} = 1$ depicts an isotropic event with particles uniformly distributed in the transverse plane. Like in our previous studies, we choose the lowest 20\% (Low-$S_{0}$) of the $S_{0}$ distribution as jetty events while the highest 20\% (High-$S_{0}$) is selected as isotropic events~\cite{Prasad:2021bdq, Mallick:2020ium}. To understand the statistical robustness of the event classes, the readers can see Ref.~\cite{Prasad:2021bdq}, where the distribution of spherocity for different centrality classes in Pb-Pb collisions is shown.

\subsection{Multi-particle cumulant method}

The anisotropic flow coefficients, $v_{\rm n}$ (of Eq.~\eqref{eq:Fourier}) are estimated using a two- and four-particle Q-cumulant method~\cite{Bilandzic:2010jr, Bilandzic:2013kga, Aamodt:2010pa}. This method gives the advantage of not requiring the information of $\psi_{\rm n}$ as well as of suppressing the non-flow effects by introducing suitable kinematic cuts. 

With $Q_{\rm n}$ vector, defined in Eq.~\eqref{eq:Qn}, the two- and four-particle azimuthal correlations are measured for each event, denoted as $\langle 2 \rangle$ and $\langle 4 \rangle$, respectively. The above-mentioned azimuthal correlations can be calculated using the following equations.
\begin{equation}
    \langle 2 \rangle = \frac{|Q_{n}|^{2} - M}{M(M-1)} 
\end{equation}
\begin{eqnarray}
    \langle 4 \rangle = \frac{|Q_{n}|^{4} + |Q_{2n}|^{2} - 2\cdot Re[Q_{2n}Q_{n}^{*}Q_{n}^{*}]}{M(M-1)(M-2)(M-3)} - \nonumber\\ 
 2\frac{2(M-2)\cdot|Q_{n}|^{2} - M(M-3)}{M(M-1)(M-2)(M-3)}
\end{eqnarray}
Here, $Q_{n}^{*}$ is the complex conjugate of $Q_{n}$, defined as follows:
\begin{equation}
    Q_{n}^{*}=\sum_{j=1}^{M}e^{-in\phi_{j}}.
\end{equation}
Further, event-average two- and four-particle azimuthal correlations, denoted as $\langle \langle 2 \rangle \rangle$ and $\langle \langle 4 \rangle \rangle$, respectively, are obtained using the following equations.
\begin{equation}
   \langle \langle 2 \rangle \rangle = \frac{\sum_{i=1}^{N_{\rm ev}} (W_{\langle 2 \rangle})_{i} \langle 2 \rangle _{i}}{\sum_{i=1}^{N_{\rm ev}} (W_{\langle 2 \rangle})_{i}}
   \label{eq:twocorr}
\end{equation}
\begin{equation}
   \langle \langle 4 \rangle \rangle = \frac{\sum_{i=1}^{N_{\rm ev}} (W_{\langle 4 \rangle})_{i} \langle 4 \rangle _{i}}{\sum_{i=1}^{N_{\rm ev}} (W_{\langle 4 \rangle})_{i}}
   \label{eq:fourcorr}
\end{equation}
Here, $N_{\rm ev}$ corresponds to the total number of events for which the event average is performed. The weight factors are estimated as, 
\begin{equation}
    W_{\langle 2 \rangle} = M(M-1),
\end{equation}
\begin{equation}
    W_{\langle 4 \rangle} = M(M-1)(M-2)(M-3).
\end{equation}
Using $\langle \langle 2 \rangle \rangle$ and $\langle \langle 4 \rangle \rangle$, we calculate the two- and four-particle cumulants using the following expressions.
\begin{equation}
c_{\rm n}\{2\} = \langle \langle 2 \rangle \rangle ,\;\;\;\;\;
c_{\rm n}\{4\} = \langle \langle 4 \rangle \rangle - 2 \langle \langle 2 \rangle \rangle ^{2}
\label{eq:cumu1}
\end{equation}
Finally, the reference flow of the particles can be estimated as follows.
\begin{equation}
v_{\rm n}\{2\} = \sqrt{c_{\rm n}\{2\}} ,\;\;\;\;  v_{\rm n}\{4\} = \sqrt[4]{- c_{\rm n}\{4\}}
\end{equation}

To make $p_{\rm T}$-differential measurement of the anisotropic flow coefficients, we start by tagging the particles as the particles of interest (POI) and the reference flow particles (RFP), which serve as the reference plane for the POIs and helps to establish the orientation of the symmetry plane. We denote the flow vector of the POIs as $p_{n}$. Similarly,  $t_{n}$ denotes the flow vector of the particles tagged as both POIs and RFPs, which are calculated using the following equations.
\begin{equation}
    p_{n} =  \sum_{j=1}^{m_{p}} e^{in\phi_{j}}
\end{equation}
\begin{equation}
    t_{n} =  \sum_{j=1}^{m_{t}} e^{in\phi_{j}}
\end{equation}
Here, $m_{p}$ refers to the total number of particles tagged as POIs, and $m_{t}$ stands for the total number of particles defined as both POIs and RFPs. Now, single-event averaged differential two- and four-particle azimuthal correlations ($\langle 2^{'} \rangle$ and $\langle 4^{'} \rangle$) are constructed as:
\begin{equation}
    \langle 2^{'} \rangle = \frac{p_{n} Q_{n}^{*} - m_{t}}{m_{p}M - m_{t}},
\end{equation}
\begin{eqnarray}
     \langle 4^{'} \rangle = [p_{n}Q_{n}Q_{n}^{*}Q_{n}^{*} - t_{2n}Q_{n}^{*}Q_{n}^{*} - p_{n}Q_{n}Q_{2n}^{*} \nonumber \\ - 2\cdot Mp_{n}Q_{n}^{*} -2\cdot m_{t}|Q_{n}|^{2} + 7\cdot t_{n}Q_{n}^{*}  \nonumber \\ -  Q_{n} t_{n}^{*} + t_{2n}Q_{2n}^{*} + 2\cdot p_{n} Q_{n}^{*} \nonumber \\ + 2\cdot m_{t}M - 6.m_{t}] \nonumber \\ / [(m_{p}M - 3m_{t})(M-1)(M-2)]
\end{eqnarray}
An event-average is performed to obtain $\langle \langle 2^{'} \rangle \rangle$ and $\langle \langle 4^{'} \rangle \rangle$ similar to Eqs.~\eqref{eq:twocorr} and \eqref{eq:fourcorr}, as follows. 
\begin{equation}
   \langle \langle 2^{'} \rangle \rangle = \frac{\sum_{i=1}^{N_{\rm ev}} (w_{\langle 2^{'} \rangle})_{i} \langle 2^{'} \rangle _{i}}{\sum_{i=1}^{N_{\rm ev}} (w_{\langle 2^{'} \rangle})_{i}}
\end{equation}
\begin{equation}
   \langle \langle 4^{'} \rangle \rangle = \frac{\sum_{i=1}^{N_{\rm ev}} (w_{\langle 4^{'} \rangle})_{i} \langle 4^{'} \rangle _{i}}{\sum_{i=1}^{N_{\rm ev}} (w_{\langle 4^{'} \rangle})_{i}} 
\end{equation}
Here, $w_{\langle 2^{'} \rangle}$ and $w_{\langle 4^{'} \rangle}$ are weight factors, calculated using the following equations.
\begin{equation}
    w_{\langle 2^{'} \rangle} = m_{p}M - m_{t}
\end{equation}
\begin{equation}
    w_{\langle 4^{'} \rangle} = (m_{p}M - 3m_{t})(M-1)(M-2)
\end{equation}
The two- and four-particle differential cumulants are then calculated as : 
\begin{equation}
d_{\rm n}\{2\} = \langle \langle 2^{'} \rangle \rangle , \\ 
d_{\rm n}\{4\} = \langle \langle 4^{'} \rangle \rangle - 2 \langle \langle 2^{'} \rangle \rangle  \langle \langle 2 \rangle \rangle .
\label{eq:cumuDiff}
\end{equation}

Finally, the two- and four-particle $p_{\rm T}$-differential flow coefficients are estimated according to :

\begin{equation}
v_{\rm n}\{2\}(p_{\rm T}) = d_{\rm n}\{2\}/ \sqrt{c_{\rm n}\{2\}}, 
\label{eq:Flow2Diff}
\end{equation}
\begin{equation}
v_{\rm n}\{4\}(p_{\rm T}) = - d_{\rm n}\{4\}/ (- c_{\rm n}\{4\})^{3/4} .
\label{eq:Flow4Diff}
\end{equation}

However, obtaining $v_{\rm n}$ via the two-particle Q-cumulant method suffers from contributions from non-flow effects~\cite{ALICE:2010suc}. Thus, we introduce a pseudorapidity gap which can substantially reduce these nonflow effects~\cite{ALICE:2011ab, Zhou:2014bba, ALICE:2014dwt, Zhou:2015iba}. With such an implementation, each event gets divided into two sub-events, $A$ and $B$, with a pseudorapidity gap, $|\Delta\eta|$, between them~\cite{Jia:2017hbm, ATLAS:2017hap, ATLAS:2017rtr}. Estimation of the two-particle correlation with pseudorapidity gap, now denoted as $\langle 2 \rangle_{\Delta\eta}$, uses flow vectors from each of the sub-events $A$ and $B$, such that:
\begin{equation}
    \langle 2 \rangle _{\Delta \eta} = \frac{Q_{n}^{A} . Q_{n}^{B^*}}{M_{A}. M_{B}}
\end{equation}
where $Q_{n}^{A}$ and $Q_{n}^{B}$ are the flow vectors corresponding to subevents $A$ and $B$, respectively. $Q_{n}^{B^*}$ is the complex conjugate of $Q_{n}^{B}$. $M_{A}$ and $M_{B}$ are the multiplicities in sub-events $A$ and $B$, respectively. Consequently, the two-particle Q-cumulant with $|\Delta\eta|$ gap reads as: 

\begin{equation}
c_{\rm n}\{2, |\Delta \eta|\} = \langle \langle 2 \rangle \rangle_{\Delta\eta}. \\
\label{eq:cumudEta}
\end{equation}

Selecting reference particles from one subevent and the POIs from the other simplifies the calculation of differential flow with the pseudorapidity gap ($\langle 2^{'} \rangle_{\Delta\eta}$) since the RFPs and POIs now belong to two disjoint sets. The single-event average differential two-particle correlation and the differential two-particle cumulant with $|\Delta\eta|$ gap, and correspondingly obtained differential flow can be estimated as follows:
\begin{equation}
    \langle 2^{'} \rangle_{\Delta\eta} = \frac{p_{n,A}Q_{n,B}^{*}}{m_{p,A}M_{B}},
\end{equation}
 
\begin{equation}
d_{\rm n}\{2, |\Delta \eta|\} = \langle \langle 2^{'} \rangle \rangle_{\Delta\eta},
\label{eq:dFloweta}
\end{equation}

\begin{equation}
v_{\rm n}\{2,|\Delta\eta|\} (p_{\rm T}) = d_{\rm n}\{2, |\Delta \eta|\}/ \sqrt{c_{\rm n}\{2, |\Delta \eta|\}}.
\label{eq:vFloweta}
\end{equation}

We use all the charged hadrons within the pseudorapidity region, $|\eta|<2.5$, to estimate anisotropic flow coefficients using the two- and four-particle Q-cumulant method. We tag the charged hadrons within $|\eta|<2.5$ and $0.2<p_{\rm T}<4.0$ GeV/c as RFPs. A pseudorapidity gap, $|\Delta\eta|>1.0$, is applied in the two-subevent method to suppress the non-flow effects from the two-particle Q-cumulant method. The statistical uncertainties are calculated using the relations shown in Ref.~\cite{Bilandzic:2012wva}. 
 
To study the relative flow fluctuations for a given flow coefficient $v_{\rm n}$, the first and second moments of the event-by-event $v_{\rm n}$ distribution are to be approximated according to~\cite{Ollitrault:2009ie, PHENIX:2018lfu}:
 
\begin{equation}
\langle v_{\rm n} \rangle \approx \sqrt{\frac{v_{\rm n}^{2}\{2,|\Delta \eta|\} + v_{\rm n}^{2}\{4\}}{2}},
\label{eq:Flowmean}
\end{equation}
\begin{equation}
\sigma_{v_{\rm n}} \approx \sqrt{\frac{v_{\rm n}^{2}\{2,|\Delta \eta|\} - v_{\rm n}^{2}\{4\}}{2}},
\label{eq:FlowSigma}
\end{equation}
It is important to note that the Eqs.~\eqref{eq:Flowmean} and \eqref{eq:FlowSigma} are applicable when the fluctuations are small and Gaussian-like~\cite{Ollitrault:2009ie}.
The relative anisotropic flow fluctuations $F(v_{\rm n})$ can be estimated as:
\begin{equation}
F(v_{\rm n}) = \frac{\sigma_{v_{\rm n}}}{\langle v_{\rm n} \rangle}.
\label{eq:FlowFluct}
\end{equation}

\section{Results and discussions}
\label{sec:results}

We start with a comparison of $v_2$ as a function of the lowest $q_2$ and ($1-S_0$) selection\footnote{We use (1-$S_0$) percentile classes instead of $S_0$ to be consistent with $q_2$ based measurement, which has a positive correlation with $v_2$.} percentiles for central, mid-central, and peripheral collisions. $q_2$ is estimated using Eq.~\eqref{eq:qn} with all kinematics similar to that of $S_0$ for a fair comparison. From the figure, it is clear that the coverage of $v_2$ with $S_0$-based event selection is broader than that of event selections with $q_2$. Not only does $S_0$ select events with higher and lower values of $v_2$ than $q_2$ in (0-10)\% centrality, but the event selection gets better as one moves towards (60-70)\% centrality class. On the other hand, the event selection with $q_2$ gets worse with lower multiplicity as the difference between the maximum and minimum values of $v_2\{2\}$ with event selection based on $q_2$ reduces from a factor of 3.8 in (0-10)\% centrality to 3.6 in (60-70)\% centrality. On the other hand, event selection based on $S_0$ selects more circular and more elliptic events in the final state than that of $q_2$.

The following sections use $S_0$ as an event shape observable to study and disentangle the contributions of $v_2$ from $v_4$ and $v_5$ in Pb-Pb collisions at $\sqrt{s_{\rm NN}}=5.02$ TeV using AMPT.

\begin{figure}
    \centering
    \includegraphics[scale=0.4]{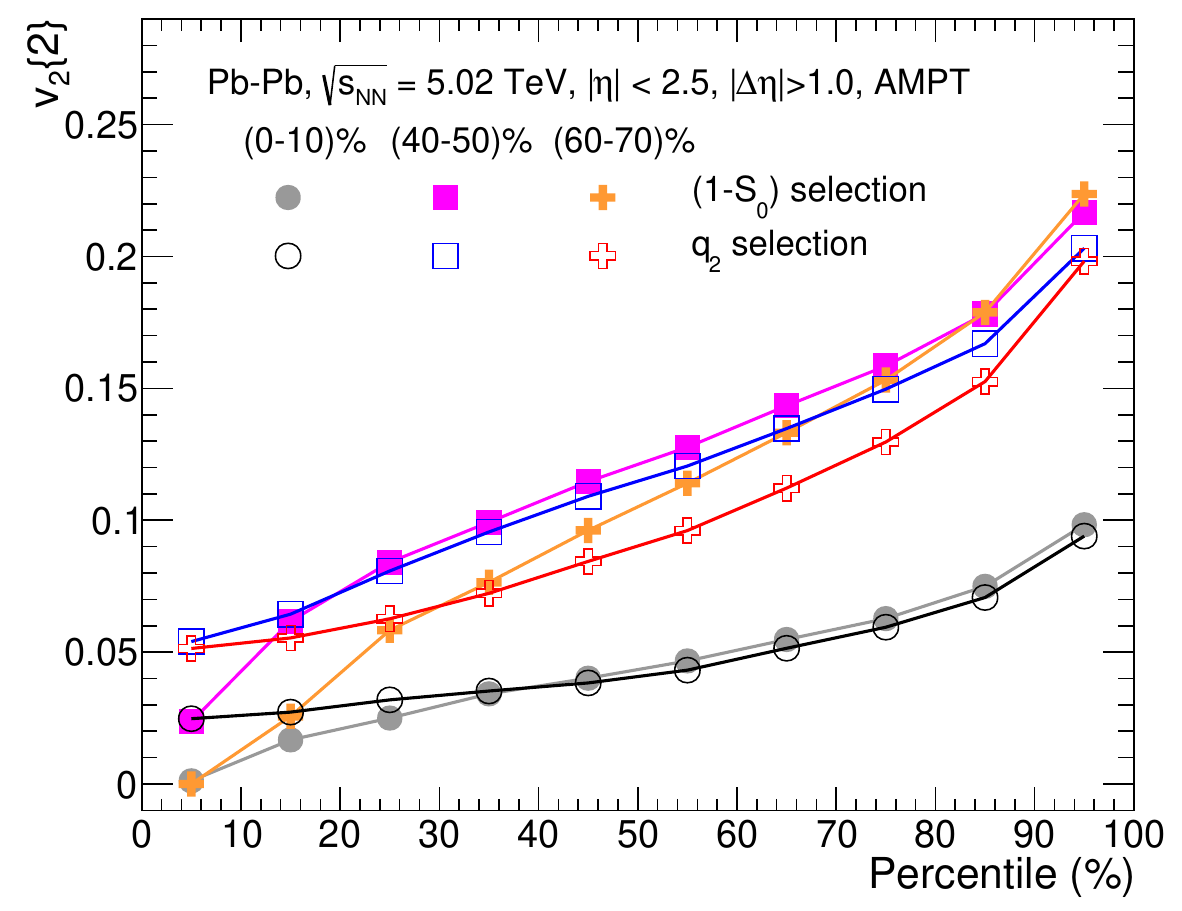}
    \caption{$v_2$ as a function of percentile classes of lowest $q_2$ and ($1-S_0$) events for different centrality classes in Pb--Pb collisions at $\sqrt{s_{\rm NN}}=5.02$ TeV using AMPT.}
    \label{fig:S0vsq2}
\end{figure}

\subsection{Spherocity dependence of elliptic flow and its fluctuations}

\begin{figure*}[ht!]
\centering
\includegraphics[scale=0.4]{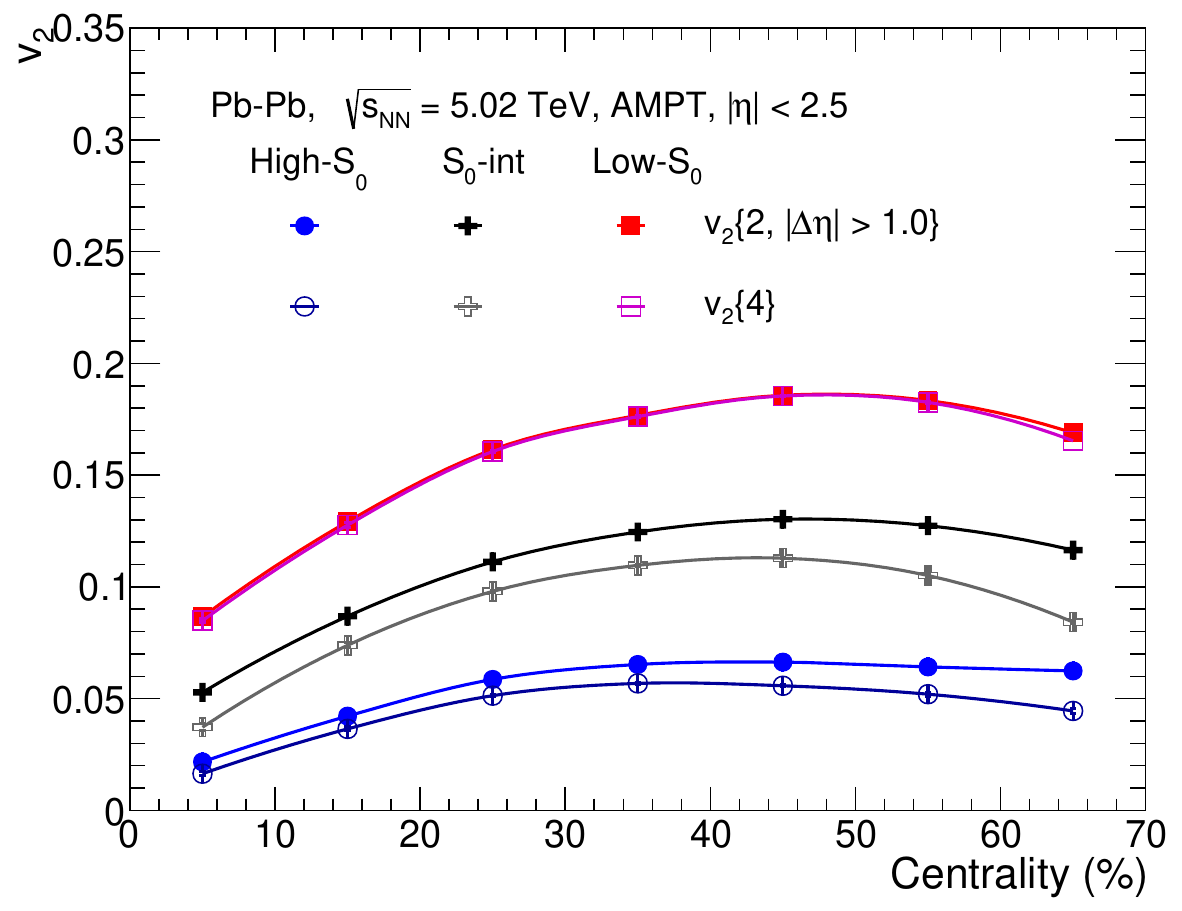}
\includegraphics[scale=0.4]{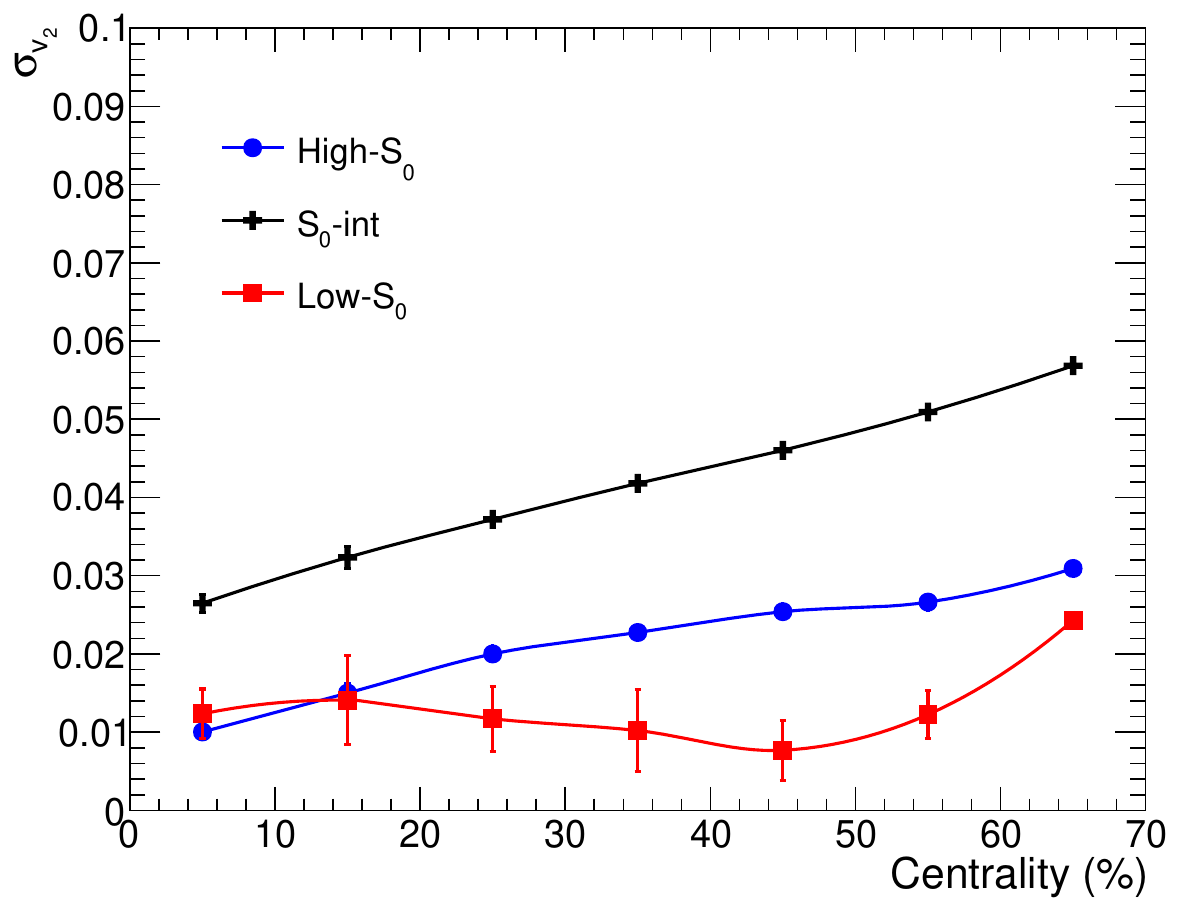}
\caption{Centrality dependence of $p_{\rm T}$-integrated $v_{2}$ (left) and $\sigma_{v_{2}}$ (right) as a function of collision centrality and transverse spherocity in Pb--Pb collisions at $\sqrt{s_{\rm NN}}=5.02$ TeV using AMPT.}
\label{fig:v2fluc}
\end{figure*}
The left panel of Fig.~\ref{fig:v2fluc} shows the centrality dependence of $v_2$ estimated using two- and four-particle Q-cumulant methods in different spherocity classes in Pb--Pb collisions at $\sqrt{s_{\rm NN}}~=~5.02$~TeV from AMPT. Here, for the $S_0$-integrated case, one finds a significant difference in the measured values of $v_{2}\{2\}$ and $v_{2}\{4\}$. This naively depicts that the $S_0$-integrated events are largely dominated by flow fluctuations due to the wide distributions of anisotropic flow coefficients when no event shape selection is applied~\cite{ATLAS:2019peb}. However, the difference between $v_{2}\{2\}$ and $v_{2}\{4\}$ decreases by selecting the high-$S_0$ events, and it becomes negligibly small for the low-$S_0$ events. Further, the difference between $v_{2}\{2\}$ and $v_{2}\{4\}$ gradually increases from central to peripheral collisions for all spherocity classes. Additionally, a saturation behavior is observed for $v_{2}\{2\}$ towards the peripheral collisions in the high-$S_0$ events. Further, $v_{2}\{4\}$ decreases towards the peripheral collisions in all event classes. This fall in $v_{2}\{4\}$ is quicker towards the peripheral collisions for the $S_0$-integrated events compared to the high- or low-$S_0$ classes. This observation suggests that the anisotropic flow fluctuations increase as the collisions become more peripheral. These observations also highlight the capability of $S_0$ to select events with smaller and larger values of $v_2$ as $S_0$ is anti-correlated with the $v_2$ of the event.~\cite{ Prasad:2022zbr, Mallick:2020ium}.

The right panel of Fig.~\ref{fig:v2fluc} depicts the centrality dependence of $v_2$ fluctuation, $\sigma_{ v_{2}}$, for different $S_0$ classes in Pb--Pb collisions at $\sqrt{s_{\rm NN}}~=~5.02$~TeV from AMPT. As expected from the results for Pb--Pb collisions at $\sqrt{s_{\rm NN}}~=~2.76$~TeV~\cite{ATLAS:2013xzf}, the value of $\sigma_{v_{2}}$ keeps on increasing from central to peripheral collisions due to the decrease in multiplicity. This trend of $\sigma_{v_{2}}$ is similar in high-$S_0$ and $S_0$-integrated events. Further, quantitatively, $\sigma_{v_{2}}$ is the largest for the $S_0$-integrated events, followed by high-$S_0$ and smallest for the low-$S_0$ events. Since $S_0$-integrated events possess events with both higher and lower values of $v_2$, the overall fluctuations are larger in this 
event class. Similarly, for high-$S_0$ events, which have dominating circular geometry in the transverse plane, their event-by-event $v_2$ can fluctuate significantly in comparison to the low-$S_0$ events where the azimuthal distribution of particles are elliptic in the ($p_{x}, p_{y}$) plane~\cite{Mallick:2020ium} and thus possess a smaller $v_2$ fluctuations. It is interesting to note that, using transverse spherocity, one can choose events with a larger $v_2$ value and also a smaller elliptic fluctuation.

\begin{figure}[ht!]
\centering
\includegraphics[scale=0.4]{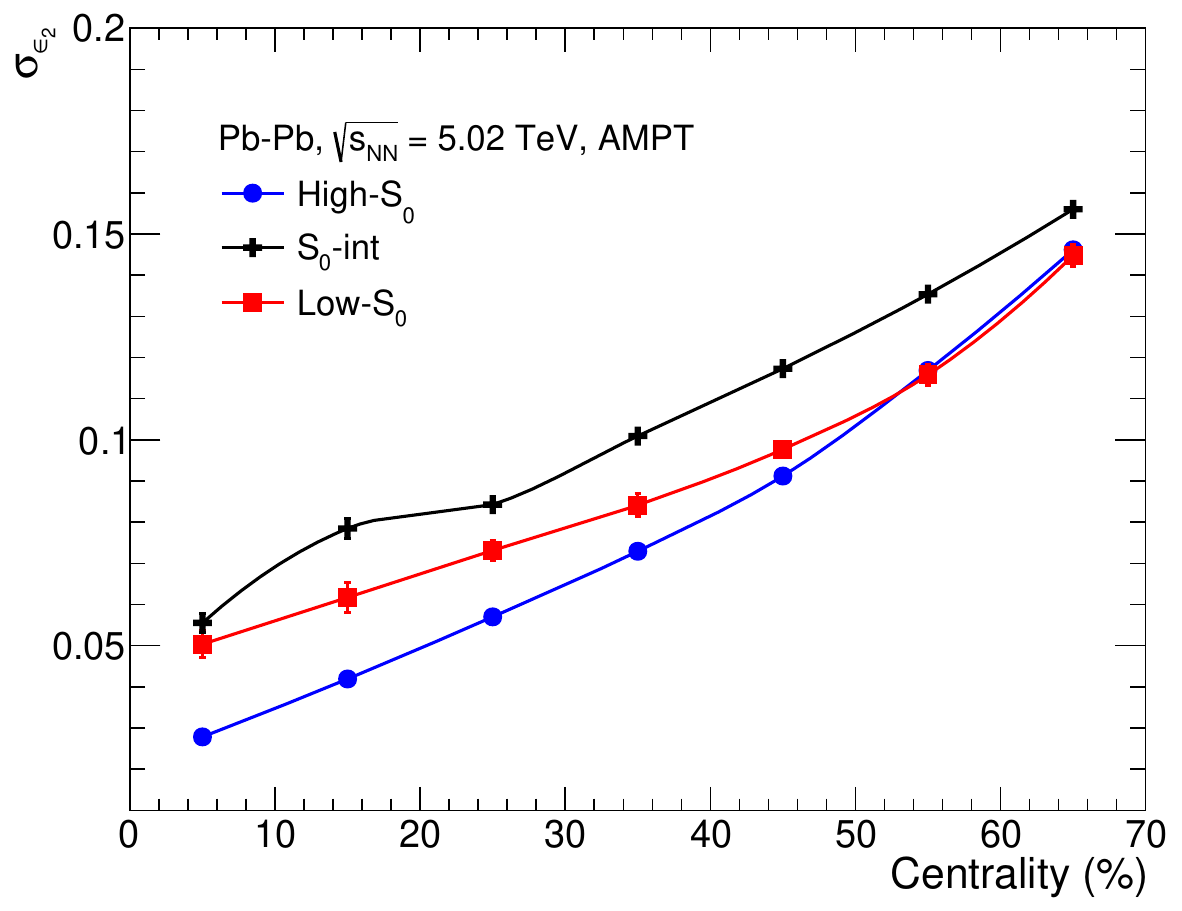}
\caption{Centrality dependence of $\sigma_{\epsilon_2}$ for high-$S_0$, $S_0$-int and low-$S_0$ events in Pb--Pb collisions at $\sqrt{s_{\rm NN}}=5.02$ TeV using AMPT.}
\label{fig:e2sigma}
\end{figure}

\begin{figure}[ht!]
\centering
\includegraphics[scale=0.4]{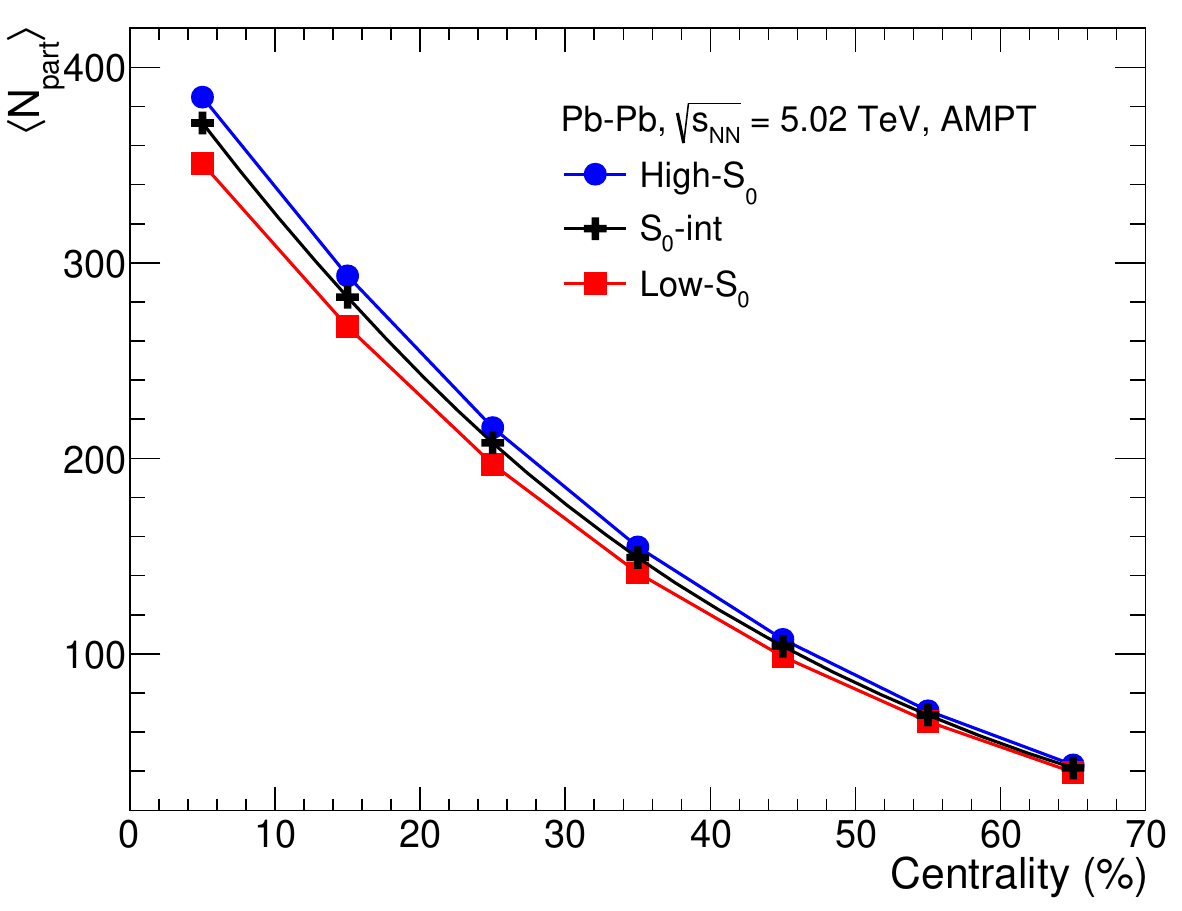}
\caption{Centrality dependence of $\langle N_{\rm part}\rangle$ for different $S_0$ classes in Pb--Pb collisions at $\sqrt{s_{\rm NN}}=5.02$ TeV using AMPT.}
\label{fig:NpartS0}
\end{figure}

Due to the probabilistic distribution of nucleons inside the colliding nuclei, event-by-event fluctuations in the collision overlap region are possible, which lead to fluctuations in the eccentricity and triangularity. These initial state fluctuations can influence the final state fluctuations and are reflected in the fluctuations of anisotropic flow coefficients convolved with the medium response during medium evolution. Studying fluctuations in the initial spatial anisotropy and the final state azimuthal anisotropy are, thus, crucial in understanding the transport properties of the medium. Hence, before analyzing the $v_2$ fluctuations for different spherocity classes, it is important to first understand the fluctuations in their initial state, \textit{i.e.}, the eccentricity fluctuations in the collision overlap region. In AMPT, eccentricity and triangularity can be calculated using the participating nucleons using the following equation~\cite{Petersen:2010cw}.

\begin{equation}
\epsilon_{n}=\frac{\sqrt{\langle{r^{n}\cos(n\phi_{\text{part}})}\rangle^{2}+\langle{r^{n}\sin(n\phi_{\text{part}})}\rangle^{2}}}{\langle{r^{n}}\rangle}
\label{eq:eccentricity} \ .
\end{equation}
Here, $r^{n}$ and $\phi_{\rm part}$ are the radial distance and azimuthal angle of the participating nucleons. $\epsilon_2$ corresponds to eccentricity, and $\epsilon_3$ would correspond to the triangularity of the collision overlap region\footnote{$\epsilon_2$ quantifies the elliptic shape of the collision overlap region and $\epsilon_3$ quantifies how much triangular the collision overlap region is.}. $\langle\dots\rangle$ in Eq.~\eqref{eq:eccentricity} indicates the average over all the participating nucleons in a given event. The average over all the events for $\epsilon_n$ is denoted as $\langle \epsilon_n\rangle$ throughout the paper. 
In Ref.~\cite{Prasad:2022zbr}, the centrality dependence of $\langle\epsilon_2\rangle$ is shown for different $S_0$ classes in Pb–Pb collisions at $\sqrt{s_{\rm NN}}=5.02$ TeV using AMPT, where we observe a rise in $\langle\epsilon_2\rangle$ from central to peripheral collisions. Further, the low-$S_0$ events have larger $\langle\epsilon_2\rangle$ and high-$S_0$ events have a smaller value of $\langle\epsilon_2\rangle$. In Fig.~\ref{fig:e2sigma}, we show $\sigma_{\epsilon_2}$ as a function of centrality and $S_0$ in Pb-Pb collisions using AMPT. Here, $\sigma_{\epsilon_2}$ is the eccentricity fluctuation, estimated as, $\sigma_{\epsilon_2}=\sqrt{\langle\epsilon_{2}^{2}\rangle-\langle\epsilon_{2}\rangle^{2}}$. From central to peripheral collisions, $\sigma_{\epsilon_2}$ increases irrespective of the $S_0$ class, which is attributed to an increase in geometry fluctuations arising due to a decrease in the number of participating nucleons, as shown in Fig.~\ref{fig:NpartS0}. $\langle N_{\rm part}\rangle$ as a function of centrality in Pb--Pb collisions at $\sqrt{s_{\rm NN}}=5.02$ TeV, shown in Fig~\ref{fig:NpartS0}, for $S_0$-int class is in agreement to the that measured in Ref.~\cite{ATLAS:2018ezv}. Further, throughout the centrality classes, $\sigma_{\epsilon_2}$ is larger for the $S_0$-int class as compared to high and low-$S_0$ classes. This is expected since for the $S_0$-int case, the distribution of $\epsilon_2$ is broader as compared to low or high-$S_0$ cases where event shape selections are applied. Moreover, low-$S_0$ has larger $\sigma_{\epsilon_2}$ as compared to high-$S_0$ in (0-50)\% centrality, and is attributed to the lower average number of participants, leading to larger geometrical fluctuations, in the low-$S_0$ events, as shown in Fig.~\ref{fig:NpartS0}. 

\begin{figure}[ht!]
\centering
\includegraphics[scale=0.4]{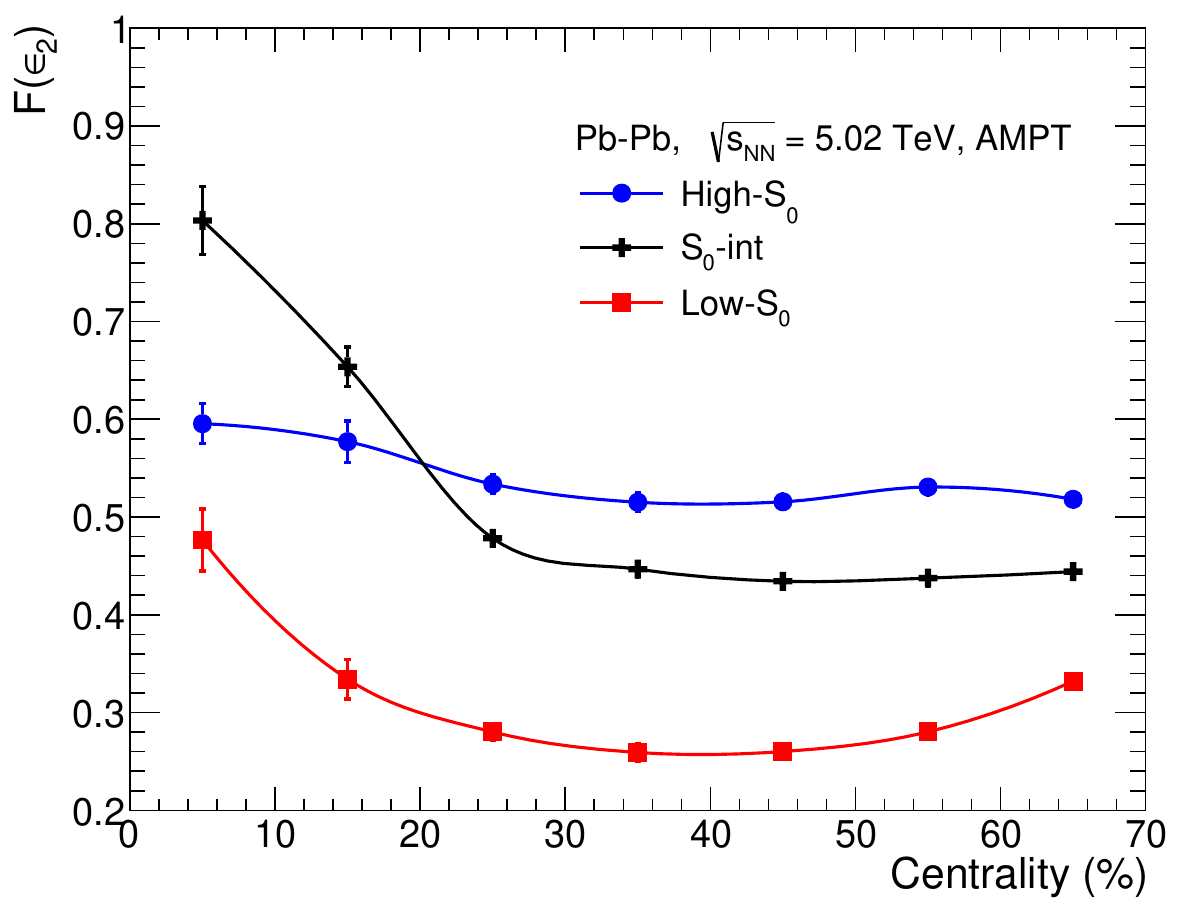}
\includegraphics[scale=0.4]{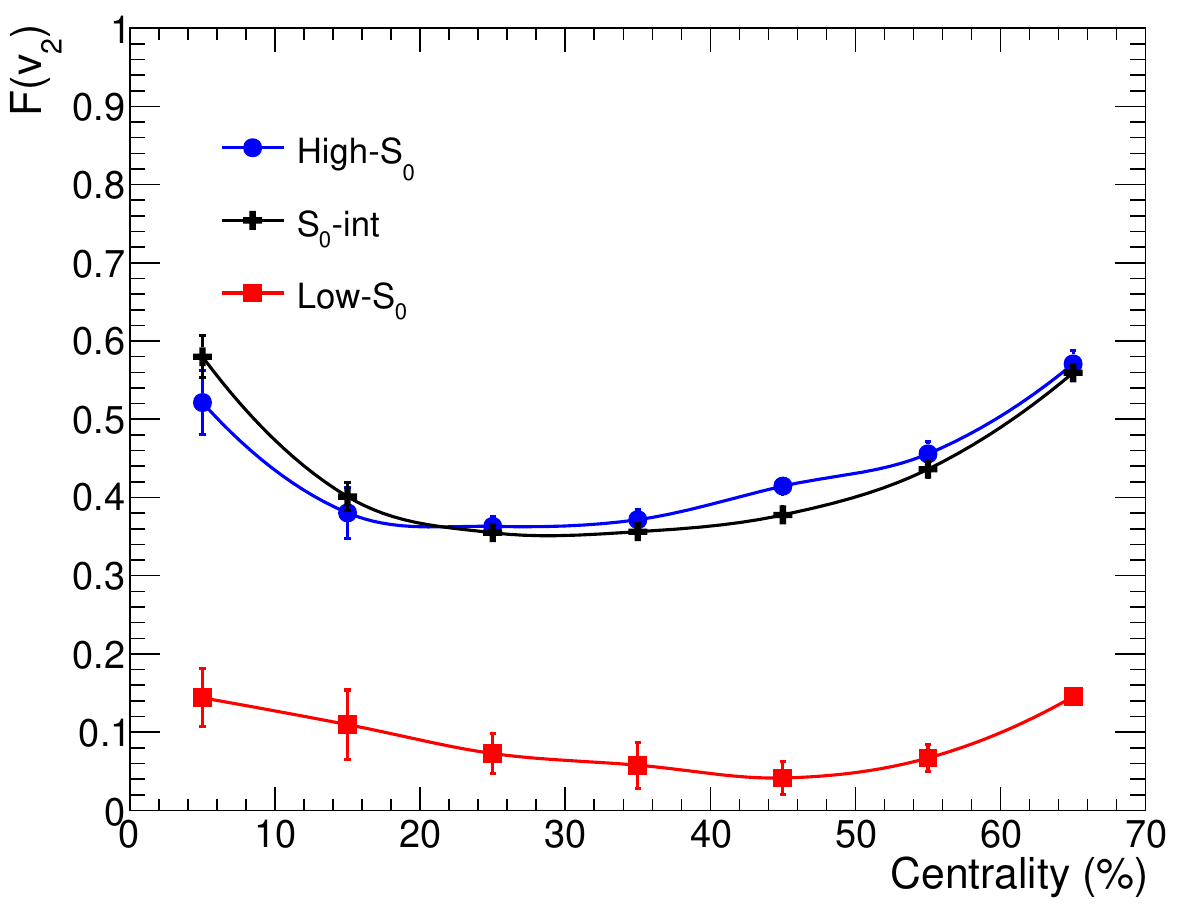}
\caption{Centrality dependence of $F(\epsilon_{2})$ (upper) and $F(v_{2})$ (lower) for different $S_0$ classes in Pb--Pb collisions at $\sqrt{s_{\rm NN}}=5.02$ TeV using AMPT.}
\label{fig:e2v2fluc}
\end{figure}

Moreover, we extend our study to show the $S_0$-dependence of relative eccentricity fluctuations ($F(\epsilon_2)$), defined as $F(\epsilon_2)=\sigma_{\epsilon_2}/\langle\epsilon_2\rangle$~\cite{Filip:2009zz, Ma:2016hkg}. The upper panel in Fig.~\ref{fig:e2v2fluc} shows the centrality dependence of eccentricity fluctuation ($F(\epsilon_{2})$) for different $S_0$ classes in Pb--Pb collisions at $\sqrt{s_{\rm NN}}=5.02$ TeV from AMPT. In each spherocity class, $F(\epsilon_{2})$ decreases from the most central to the mid-central bin. Thereafter, $F(\epsilon_{2})$ remains almost flat for $S_0$-integrated and high-$S_0$ events from mid-central to peripheral collisions, while for low-$S_0$, $F(\epsilon_{2})$ begins to rise towards the peripheral collisions. In the (0-20)\% centrality, $S_0$-integrated events have the largest relative eccentricity fluctuations, while the low-$S_0$ events have the least fluctuations throughout the centrality classes. However, the value of $F(\epsilon_2)$ for the $S_0$-int class falls between the high and low-$S_0$ classes in the mid-central and peripheral collisions. This behaviour can be understood as follows. In the most central collisions, the measured values of $\epsilon_2$ are driven significantly by event-by-event density fluctuations. A large value of $\sigma_{\epsilon_2}$ and small $\langle\epsilon_2\rangle$ for the $S_0$-int case in central collisions lead to a large $F(\epsilon_2)$. In contrast, a large value of $\langle\epsilon_2\rangle$ for the low-$S_0$ events makes $F(\epsilon_2)$ small. However, towards the mid-central collisions, the difference in the values of $\sigma_{\epsilon_2}$  for different $S_0$ classes is comparatively smaller than that of $\langle\epsilon_2\rangle$, which drives the values of $F(\epsilon_2)$ in the mid-central or peripheral collisions, shown in Fig.~\ref{fig:e2v2fluc}.
Quantitatively, the high-$S_0$ and low-$S_0$ events have a significant and distinct separation in their value of $F(\epsilon_{2})$ across all the centrality bins under study. This suggests that the transverse spherocity can also disentangle events based on the degree of fluctuations in the initial geometry. 

\begin{figure*}[ht!]
\centering
\includegraphics[scale=0.29]{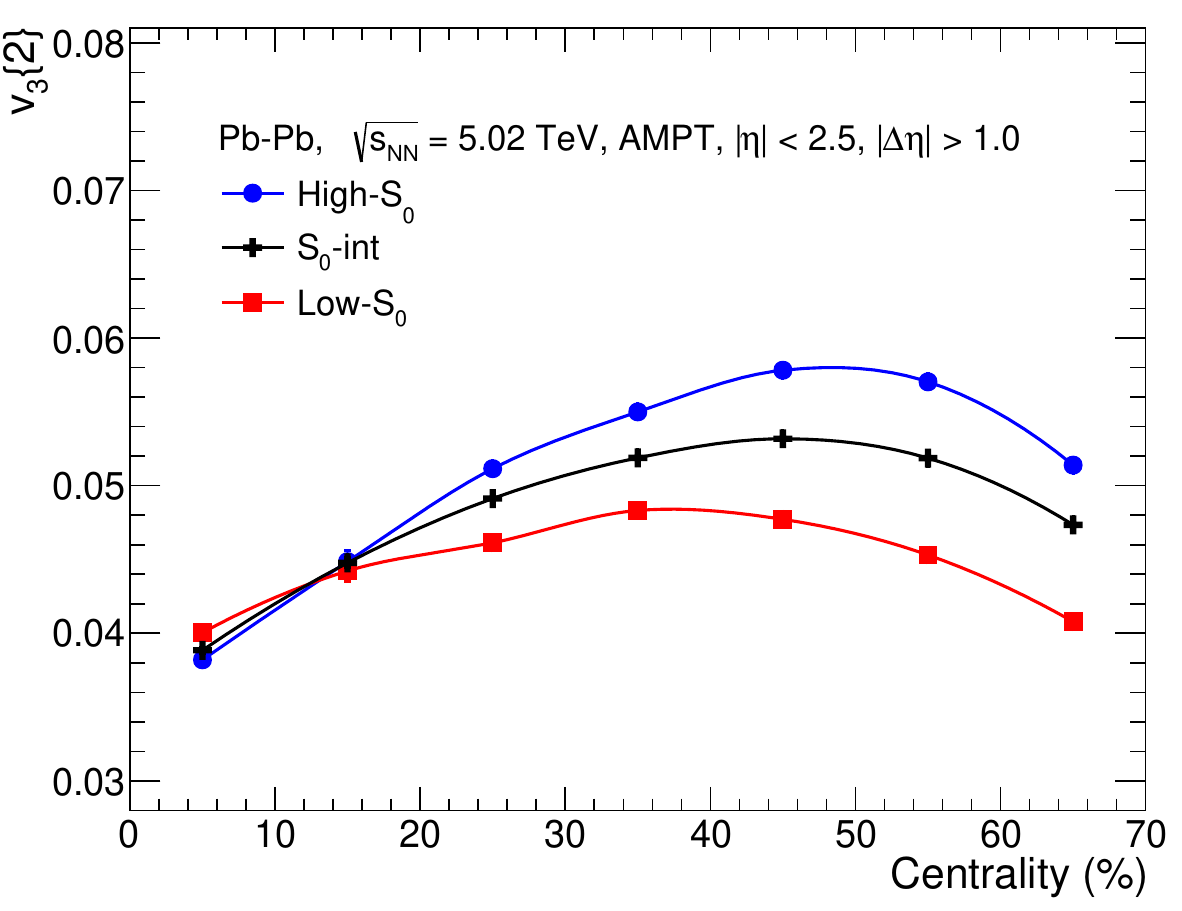}
\includegraphics[scale=0.29]{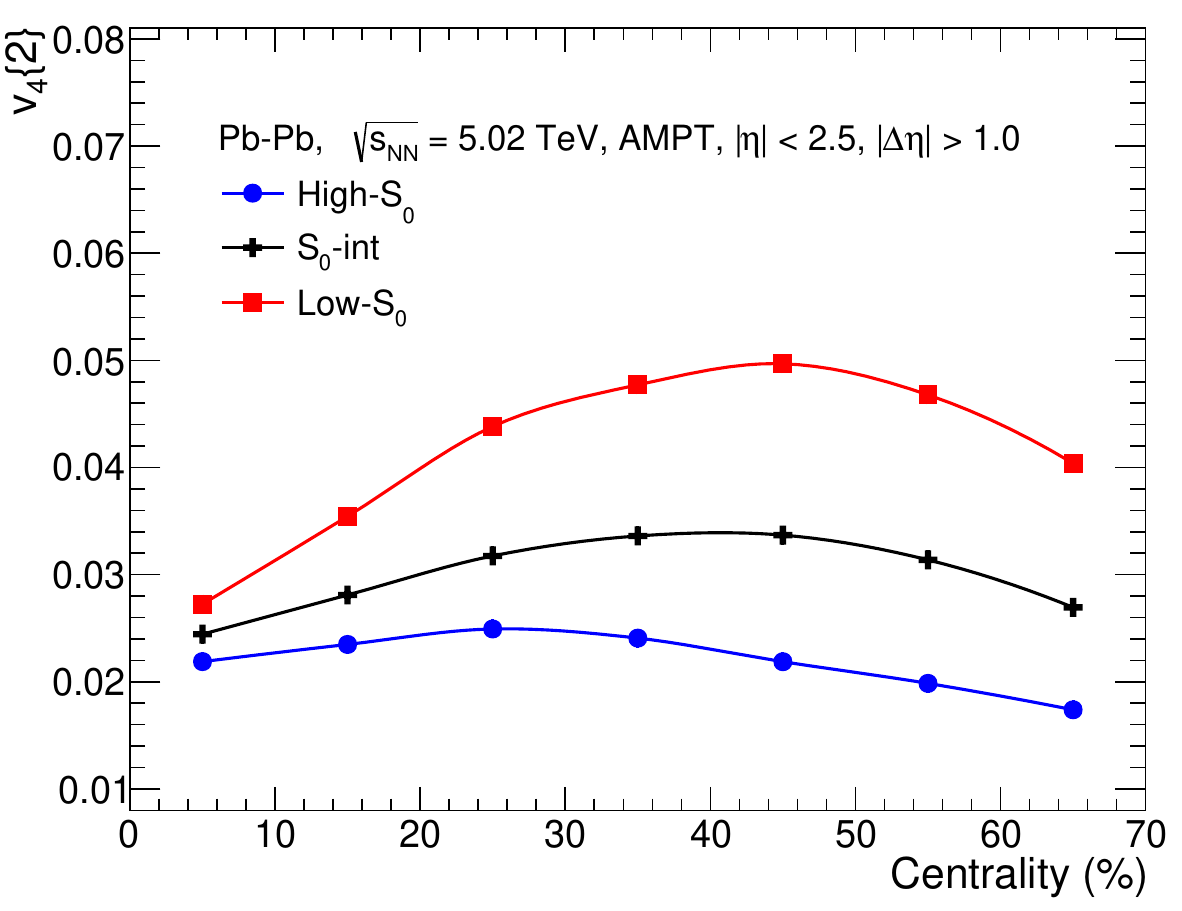}
\includegraphics[scale=0.29]{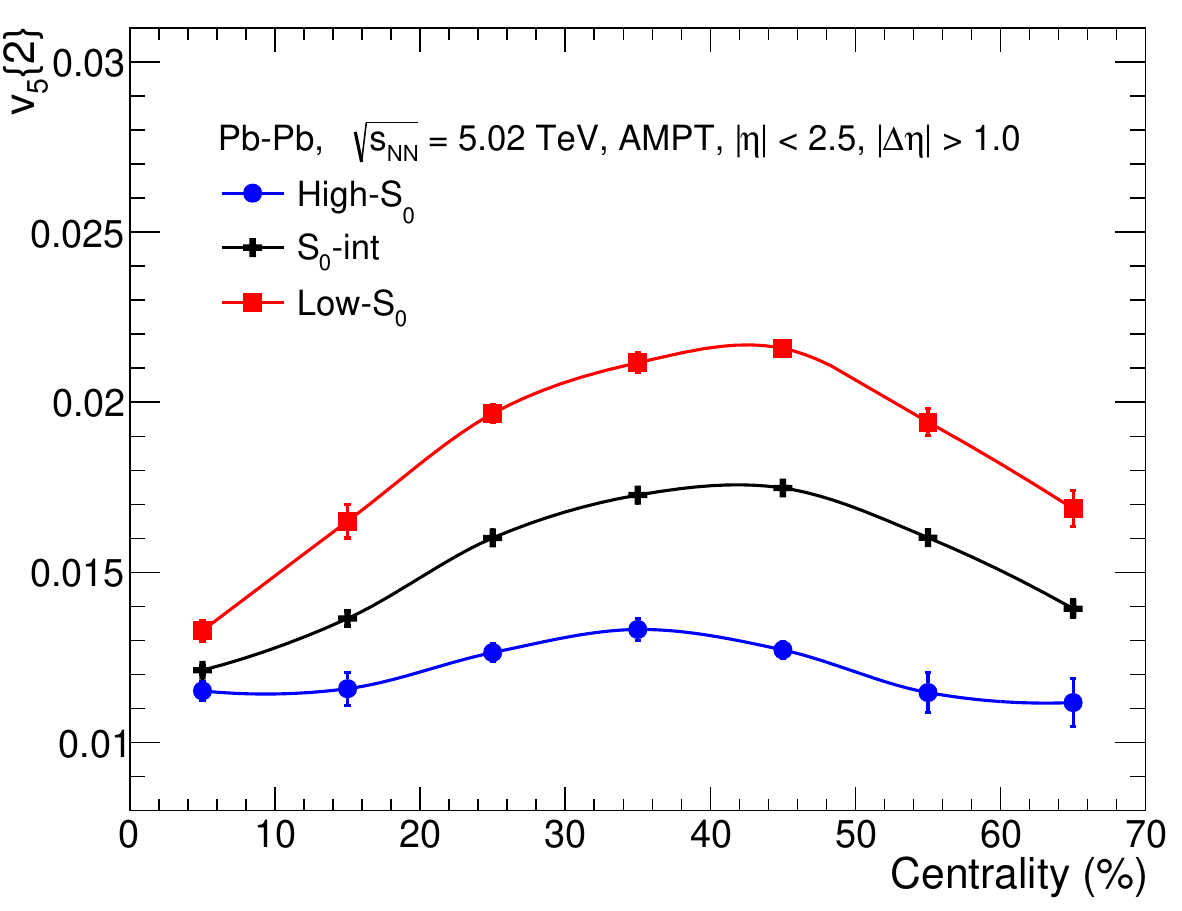}
\caption{Centrality dependence of $p_{\rm T}$-integrated $v_{3}\{2\}$ (left), $v_{4}\{2\}$ (middle) and $v_{5}\{2\}$ (right) in for different $S_0$ classes in Pb--Pb collisions at $\sqrt{s_{\rm NN}}=5.02$ TeV using AMPT.}
\label{fig:vnvscent}
\end{figure*}

In the lower panel of Fig.~\ref{fig:e2v2fluc}, we show the relative $v_2$ fluctuations, $F(v_{2})=\sigma_{v_{2}}/\langle v_{2}\rangle$, as a function of collision centrality for $S_0$-integrated, high-$S_0$ and low-$S_0$ events in Pb--Pb collisions at $\sqrt{s_{\rm NN}}~=~5.02$~TeV from AMPT. Here, $F(v_{2})$ is found to be minimum for the mid-central collisions while possessing larger values in the most-central and peripheral collisions, which can also be inferred from the left panel of Fig.~\ref{fig:v2fluc}, where the difference between $v_{2}\{2\}$ and $v_{2}\{4\}$ becomes minimum in the mid-central case. This observation is consistent with the results for Pb--Pb collisions at $\sqrt{s_{\rm NN}}~=~2.76$~TeV as shown in Refs.~\cite{ATLAS:2013xzf, ATLAS:2014qxy, CMS:2013wjq} by ATLAS(CMS) Collaboration, where the relative flow fluctuations are measured as a function of $N_{\rm part}$(centrality). The large value of $F(v_{2})$ in the most central collisions naively indicates that $v_2$ measured in the central collisions has significant contributions from the flow fluctuations, which result from large initial eccentricity fluctuations. Further, towards the peripheral collisions, although the nuclear overlap region is more elliptic in the transverse plane, a smaller number of final state particle multiplicity makes the fluctuations grow in this region. However, the mid-central collisions have both an elliptical nuclear overlap region and a large number of particle yields, which together lead to a smaller value of $F(v_{2})$ in the mid-central collisions. A significant $S_0$ dependence is also observed in $F(v_{2})$. Here, the low-$S_0$ events that have a larger $v_2$ contribution are found to possess the lowest value of $F(v_{2})$. In addition, the values of $F(v_{2})$ for the $S_0$-integrated and high-$S_0$ events are almost overlapping with each other, where high-$S_0$ events lead $F(v_{2})$ by a slightly higher value, beyond central collisions. These observations of transverse spherocity dependence of $F(v_{2})$ are found to have propagated significantly from the initial state effects, as shown in the upper plot of Fig.~\ref{fig:e2v2fluc} for $F(\epsilon_{2})$. Therefore, one can select events with smaller $v_2$ fluctuations by choosing the low-$S_0$ event classes.
\begin{figure*}[ht!]
\centering
\includegraphics[scale=0.4]{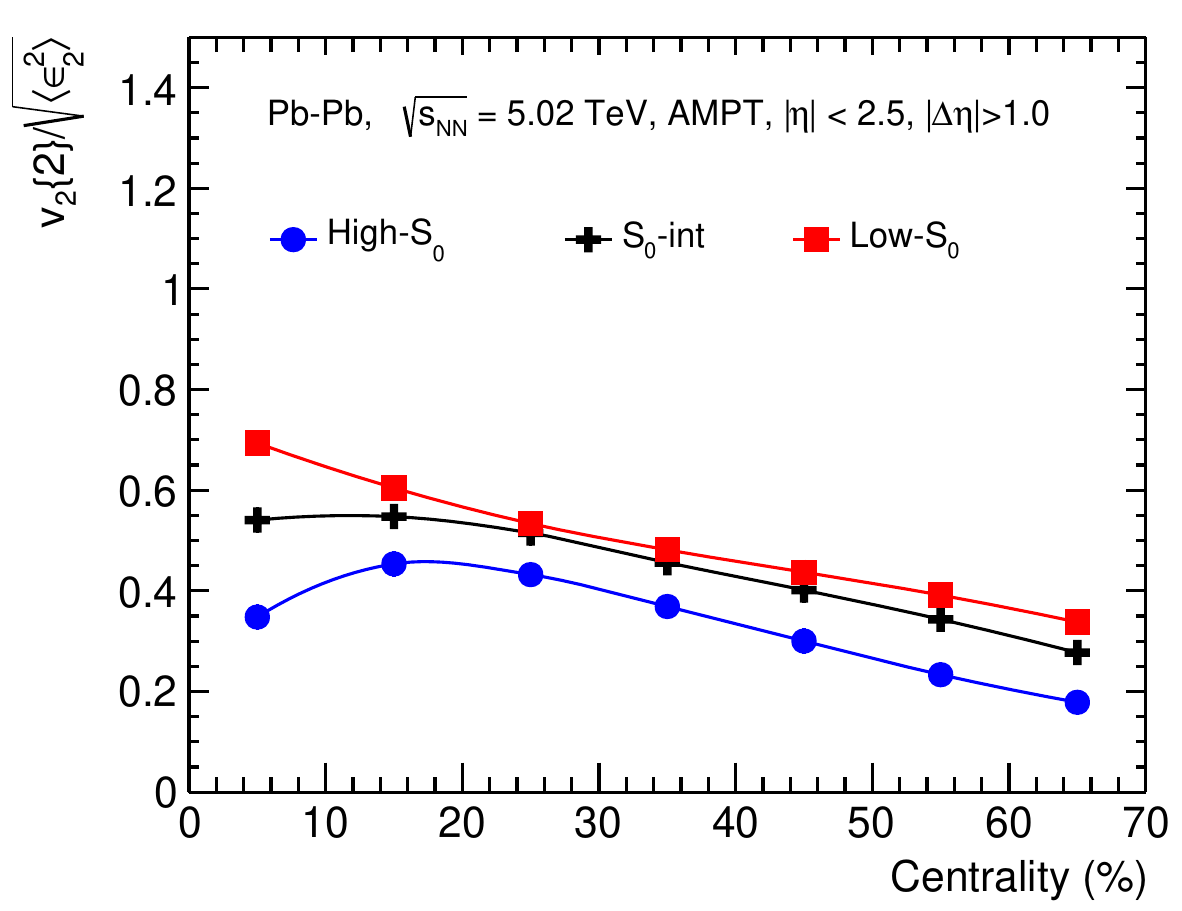}
\includegraphics[scale=0.4]{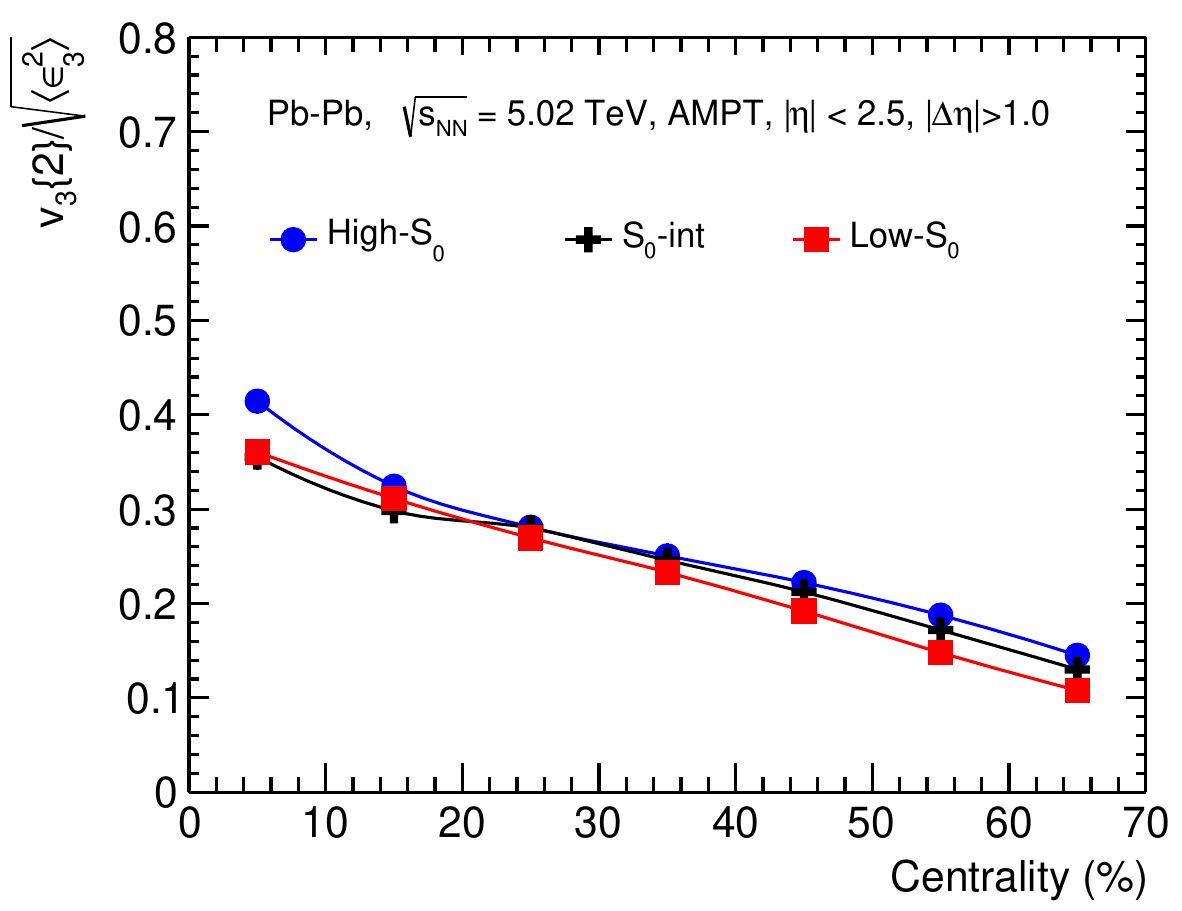}
\includegraphics[scale=0.4]{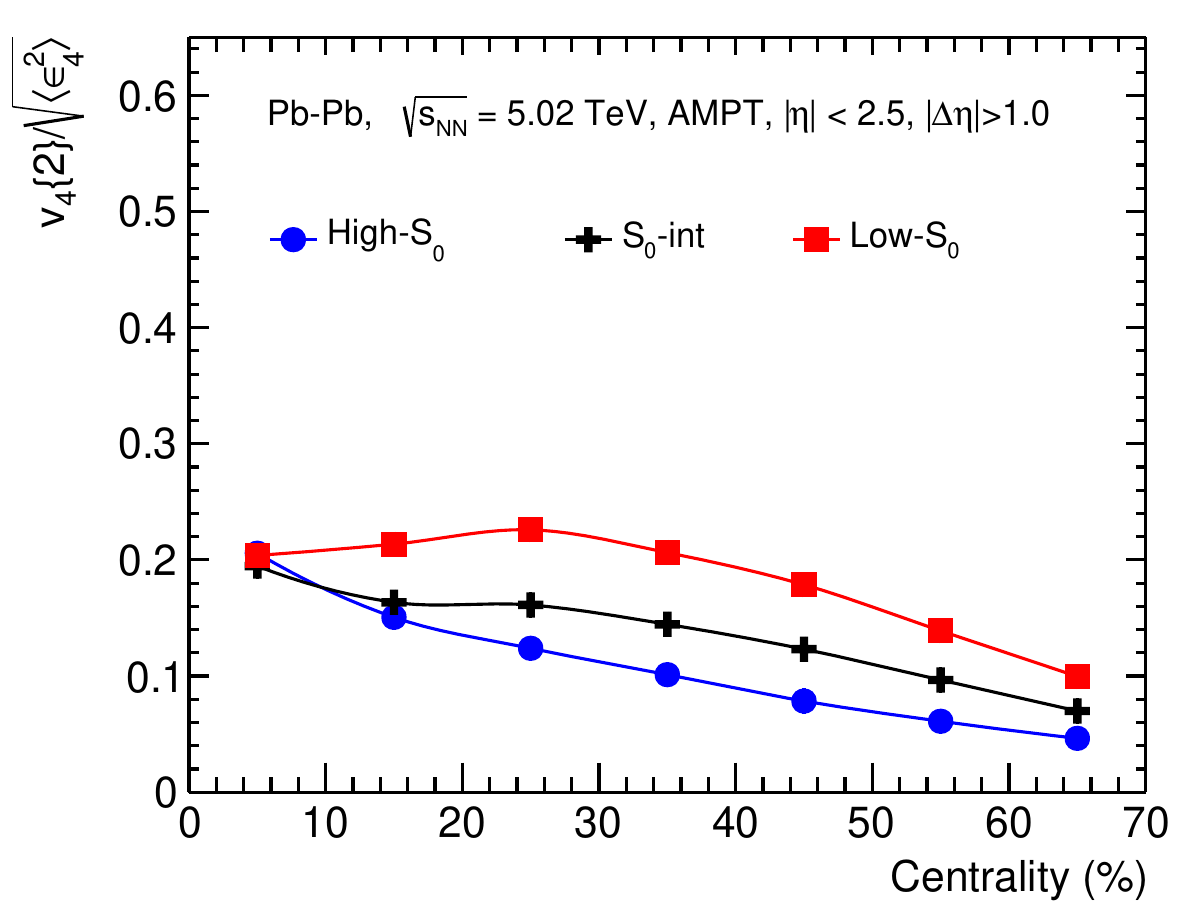}
\includegraphics[scale=0.4]{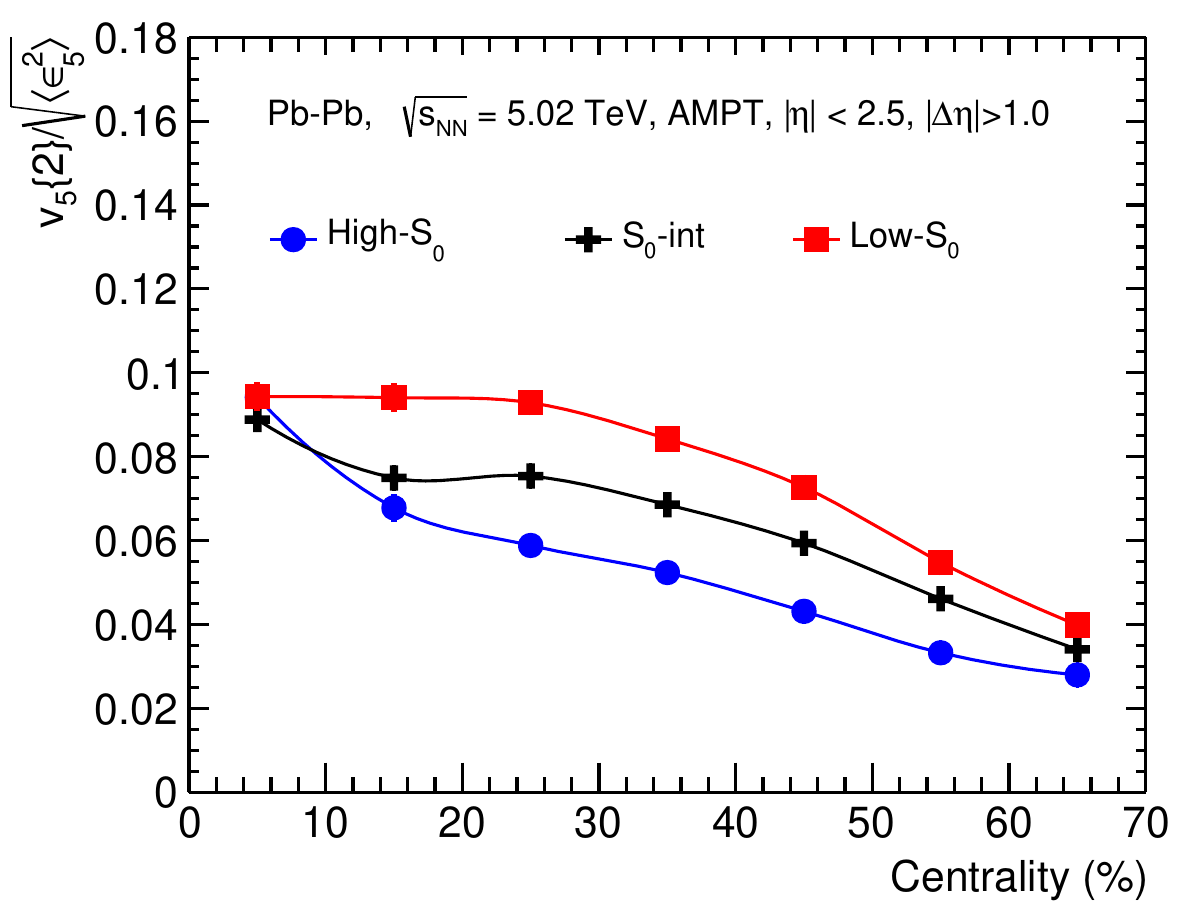}
\caption{Centrality dependence of $v_{\rm n}\{2\}/\sqrt{\langle \epsilon_{\rm n}^{2}\rangle}$, with n=2, 3, 4, and 5, for different $S_0$ classes in Pb-Pb collisions at $\sqrt{s_{\rm NN}}=5.02$ TeV using AMPT.}
\label{fig:vnenratios}
\end{figure*}

\subsection{Spherocity dependence of 3rd and higher order anisotropic flow coefficients}
Figure~\ref{fig:vnvscent} shows $p_{\rm T}$-integrated $v_{3}\{2\}$ (left), $v_{4}\{2\}$ (middle) and $v_{5}\{2\}$ (right) of all charged hadrons as a function of collision centrality in Pb--Pb collisions at $\sqrt{s_{\rm NN}}=5.02$ TeV in different $S_0$ classes from AMPT. For the $S_0$-integrated case, $v_{n}$ gradually increases up to the mid-central collisions to attain a peak, and then starts to fall smoothly towards the peripheral collisions. This is because the collision overlap region in central collisions is fairly isotropic in the $xy$-plane, which develops smaller values of azimuthal anisotropy in the final state. However, as one moves towards the mid-central collisions, the collision overlap region begins to get more elliptic in shape, meaning it is no longer isotropic in the $xy$-plane, leading to enhanced values of $v_{n}\{2\}$. In addition to that, the number of participants also decreases towards peripheral collisions, as a result of which, density fluctuations become stronger. Again, towards the peripheral collisions, the smaller number of participants and smaller lifetime of the fireball constrain the development of the initial anisotropies to the final state. This leads to the reduction of $v_{n}\{2\}$ values towards peripheral collision. In addition, one observes from Fig.~\ref{fig:v2fluc} and Fig.~\ref{fig:vnvscent} that $v_{2}>v_{3}>v_{4}>v_{5}$ ordering is well respected throughout the collision centralities for the $S_{0}$-integrated events, meaning, $v_2$ has the largest contribution to azimuthal anisotropy in heavy-ion collisions, and the contributions decrease from lower to higher order harmonics. Interestingly, the values of $v_{n}$, except for $n=3$, are found to be enhanced for the events having low-$S_0$ values and suppressed for high-$S_0$ events in comparison to the $S_0$-integrated events. Contrary to this, $v_3$ is observed to be enhanced for the high-$S_0$ events, while low-$S_0$ events show smaller values of $v_{3}\{2\}$. The $S_0$ class dependence of $v_2$ and $v_3$ is consistent with previous measurements shown using the two-particle correlation method~\cite{Prasad:2022zbr}. Here, it is also shown that transverse spherocity is (anti)correlated to ($q_{2}$)$q_{3}$- based event selection. This makes the transverse spherocity dependent features of $v_n$ similar to the results from $q_{2}$-based event selection~\cite{ATLAS:2015qwl}. For instance, $v_{2}\{2\}$, $v_{4}\{2\}$ and $v_{5}\{2\}$ are positively correlated with $q_{2}$ which makes them anti-correlated with event selections based on $S_0$. Similarly, since $v_{3}$ and $v_{2}$ are anti-correlated, $v_{3}\{2\}$ shows a positive correlation with $S_{0}$~\cite{ATLAS:2015qwl}. As discussed earlier in Section~\ref{sec:intro}, $v_{4}\{2\}$ has non-linear contributions from $v_2$. As a consequence, the anti-correlation of $v_{2}\{2\}$ with $S_0$ is reflected in $v_{4}\{2\}$. Likewise, $v_{5}\{2\}$ has contributions from both $v_{2}\{2\}$ and $v_{3}\{2\}$, which is shown in Eq.~\eqref{eq:v5vsv2v3}. Since, throughout the central classes, $v_{2}$ dominates over $v_3$, one would expect $v_5$ to have a larger contribution from $v_2$ than $v_3$. Since the anti-correlation between $v_{2}\{2\}$ and $S_0$ stronger than the correlation between $v_{3}\{2\}$ and $S_0$, one observes anti-correlation between $v_{5}\{2\}$ and $S_0$.

\subsection{Spherocity dependence of $v_{n}\{2\}/\sqrt{\langle\epsilon_n^2\rangle}$}

One of the ways to characterize the medium response to the evolution of final state azimuthal anisotropy from the initial spatial anisotropy is the ratio $v_{n}/\epsilon_n$. Figure~\ref{fig:vnenratios} shows the centrality dependence of $v_{n}\{2\}/\sqrt{\langle\epsilon_n^2\rangle}$ in Pb--Pb collisions at $\sqrt{s_{\rm NN}}=5.02$ TeV in various $S_0$ classes from AMPT. In the upper left panel, one finds a significant $S_0$ and centrality dependence on $v_{2}\{2\}/\sqrt{\langle\epsilon_2^2\rangle}$. The ratio $v_{2}\{2\}/\sqrt{\langle\epsilon_2^2\rangle}$ decreases from central to peripheral collisions, which could be attributed to the reduced number of participants, indicating a strong impact of the system size on the evolution of $\epsilon_2$ to $v_2$. $v_{2}\{2\}/\sqrt{\langle\epsilon_2^2\rangle}$ is found to be higher for the low-$S_0$ events while, this ratio is smaller for the high-$S_0$ events. Since high-$S_0$ events have larger final state multiplicity as compared to low-$S_0$ events~\cite{Prasad:2021bdq}, one would expect $v_{2}\{2\}/\sqrt{\langle\epsilon_2^2\rangle}$ to be higher for high-$S_0$ events for a particular centrality class. Therefore, the unexpected results of $v_{2}\{2\}/\sqrt{\langle\epsilon_2^2\rangle}$ could be understood through the following possible scenarios. 
\begin{itemize}
    \item Firstly, in the high-$S_0$ events, the large number of soft partonic interactions drives the medium towards isotropization in the final state; hence, high-$S_0$ events tend to adversely affect the transformation of $\epsilon_2$ to $v_2$.
    \item This also means that the system response to the geometrical anisotropy differs from that of the anisotropy arising out of initial density fluctuations, as one observes a smaller $v_{2}\{2\}/\sqrt{\langle\epsilon_2^2\rangle}$ where the event-by-event fluctuations dominate the contribution of $v_2$. 
    \item Further, a decrease of $v_{2}\{2\}/\sqrt{\langle\epsilon_2^2\rangle}$ in the most central collisions of high-$S_0$ events also hints that the events, where the $v_2$ fluctuations dominate, have higher sensitivity to the system evolution.
    
\end{itemize}
In the upper right panel of Fig.~\ref{fig:vnenratios}, $v_{3}\{2\}/\sqrt{\langle\epsilon_3^2\rangle}$ is shown as a function of centrality and spherocity. Similar to $v_{2}\{2\}/\sqrt{\langle\epsilon_2^2\rangle}$, a decrease in $v_{3}\{2\}/\sqrt{\langle\epsilon_3^2\rangle}$ is observed from central to peripheral collisions. However, $v_{3}\{2\}/\sqrt{\langle\epsilon_3^2\rangle}$ show negligible spherocity dependence~\cite{Prasad:2022zbr}. In (40-70)\% centrality class, where $v_2$ is large, low-$S_0$ events are found to have slight smaller value of $v_{3}\{2\}/\sqrt{\langle\epsilon_3^2\rangle}$ and high-$S_0$ events possesses slightly higher value as compared to $S_0$-int events. This is expected because, in Ref.~\cite{Prasad:2022zbr}, it is shown that $\epsilon_3$ has no effect on event selection based on $S_0$, while we observe finite $S_0$ dependence in $v_3\{2\}$ in Fig.~\ref{fig:vnvscent}, which arises due to anti-correlation between $v_2$ and $v_3$. In the lower panels we show the variation of $v_{4}\{2\}/\sqrt{\langle\epsilon_4^2\rangle}$ (left) and $v_{5}\{2\}/\sqrt{\langle\epsilon_5^2\rangle}$ (right) with $S_0$ and centrality. Interestingly, both $v_{4}\{2\}/\sqrt{\langle\epsilon_4^2\rangle}$ and $v_{5}\{2\}/\sqrt{\langle\epsilon_5^2\rangle}$ show similar variation with $S_0$ and centrality selections. As expected, $v_{5}\{2\}/\sqrt{\langle\epsilon_5^2\rangle}$ has smaller values as compared to $v_{4}\{2\}/\sqrt{\langle\epsilon_4^2\rangle}$ due to a higher sensitivity to system response. $v_{n}\{2\}/\sqrt{\langle\epsilon_n^2\rangle}$ for $n=4,5$ have higher values for the low-$S_0$ events, which have large $v_2$ values. Analogously, high-$S_0$ events with smaller $v_2$ values show diminished $v_{4}\{2\}/\sqrt{\langle\epsilon_4^2\rangle}$ and $v_{5}\{2\}/\sqrt{\langle\epsilon_5^2\rangle}$. For high-$S_0$ events, since the contribution of $v_2$ is reduced, one observes a monotonic decreasing trend from central to peripheral collisions. Further high-$S_0$ events show a stronger centrality dependence of $v_{4}\{2\}/\sqrt{\langle\epsilon_4^2\rangle}$ and $v_{5}\{2\}/\sqrt{\langle\epsilon_5^2\rangle}$. This behaviour of $v_{4}\{2\}/\sqrt{\langle\epsilon_4^2\rangle}$ and $v_{5}\{2\}/\sqrt{\langle\epsilon_5^2\rangle}$ for high-$S_0$ events is similar to stronger viscous-damping effects, shown in hydrodynamic calculations~\cite{Teaney:2010vd, Alver:2010dn, Schenke:2011bn}. Moreover, $v_{4}\{2\}/\sqrt{\langle\epsilon_4^2\rangle}$ for the low-$S_0$ events shows a small rise in (20-30)\% central class, which can be attributed to the competing effects of damping (similar to high-$S_0$ events) and dominating effects from $v_2$. Since, $v_{5}\{2\}/\sqrt{\langle\epsilon_5^2\rangle}$ shows a stronger damping effect than $v_{4}\{2\}/\sqrt{\langle\epsilon_4^2\rangle}$, and possibly possesses a weaker contribution from $v_2$, a saturation behaviour is observed for $v_{5}\{2\}/\sqrt{\langle\epsilon_5^2\rangle}$ in (0-30)\% centrality class for low-$S_0$ events. The centrality dependence of $v_{n}\{2\}/\sqrt{\langle\epsilon_n^2\rangle}$ we obtain for Pb--Pb collisions at $\sqrt{s_{\rm NN}}=5.02$ TeV, has a decent qualitative agreement with the $N_{\rm part}$-dependence of $v_{n}\{2\}/\sqrt{\langle\epsilon_n^2\rangle}$ observed for Pb--Pb collisions at $\sqrt{s_{\rm NN}}=2.76$ TeV, with the differences in magnitude owed to the difference in the collision energies compared~\cite{ATLAS:2015qwl}.

\section{Summary}
\label{sec:summary}

In summary, we have studied the transverse spherocity dependence of anisotropic flow coefficients with a focus on the effects of $v_2$ on higher-order flow coefficients and $v_2$ fluctuations in Pb--Pb collisions at $\sqrt{s_{\rm NN}}=5.02$ TeV using AMPT. We observe that, except for triangular flow, which has a positive correlation with transverse spherocity, higher order flow coefficients, such as $v_4$ and $v_5$, show an anti-correlation with transverse spherocity. This is expected as higher-order coefficients have large non-linear contributions from $v_2$, which is picked up and can be removed with proper transverse spherocity selections. Further, one finds that the event selections based on transverse spherocity can affect the $v_2$ fluctuations. The $v_2$ fluctuations are the smallest for the low-$S_0$ events, where the signal of $v_2$ is stronger. Further, with the help of the event shape classifier, we find that the system response to the $v_2$ coefficient from geometry is different from the fluctuation.  
\\

The present study encourages the use of transverse spherocity over the reduced flow vectors for the event-shape-based studies of anisotropic flow coefficients in heavy-ion collisions. This is because transverse spherocity-based event selection has a higher coverage on the values of $v_2$ as compared to that of reduced flow vectors from most central to peripheral collisions. This can be easily cross-checked in current experiments at the LHC, with similar measurements shown in Fig.~\ref{fig:S0vsq2} of the paper. Moreover, the present study shows that with event-shape selection based on transverse spherocity, one can select events with extreme values of $v_2$ (larger or smaller), and also simultaneously select events with extreme values of $F(v_{2})$, respectively. This unique feature of transverse spherocity can be exploited in experiments to select events with small $v_{2}$ so as to reduce the contribution of $v_{2}$ from higher-order flow coefficients. Further, events with the least contribution of $v_{2}$ are found to have stronger damping effects in $v_{4}$ and $v_{5}$ with the change in collision centrality. This change in the damping strength of higher-order flow coefficients with a change in centrality and spherocity class can be explored in experiments with input from different MC event generators that can accurately calculate the initial eccentricities of the collision overlap region. The method presented in this paper would thus be useful to understand the transport properties of the medium through the studies of higher-order harmonics in heavy-ion collisions. The applicability of this method in low-multiplicity events would also be helpful to explore small system dynamics ranging from pp, p-Pb to O-O and Ne-Ne collisions planned at the LHC.

\section*{Acknowledgement}
S.P. acknowledges the doctoral fellowship from the University Grants Commission (UGC), Government of India. A.M.K.R. acknowledges the doctoral fellowships from the DST INSPIRE program of the Government of India. The authors gratefully acknowledge the DAE-DST, Government of India, funding under the mega-science project “Indian participation in the ALICE experiment at CERN” bearing Project No. SR/MF/PS-02/2021-IITI(E-37123). N.M. is supported by the Academy of Finland through the Center of Excellence in Quark Matter with Grant No. 346328.

\section*{Appendix}
\begin{figure}[ht!]
\centering
\includegraphics[scale=0.4]{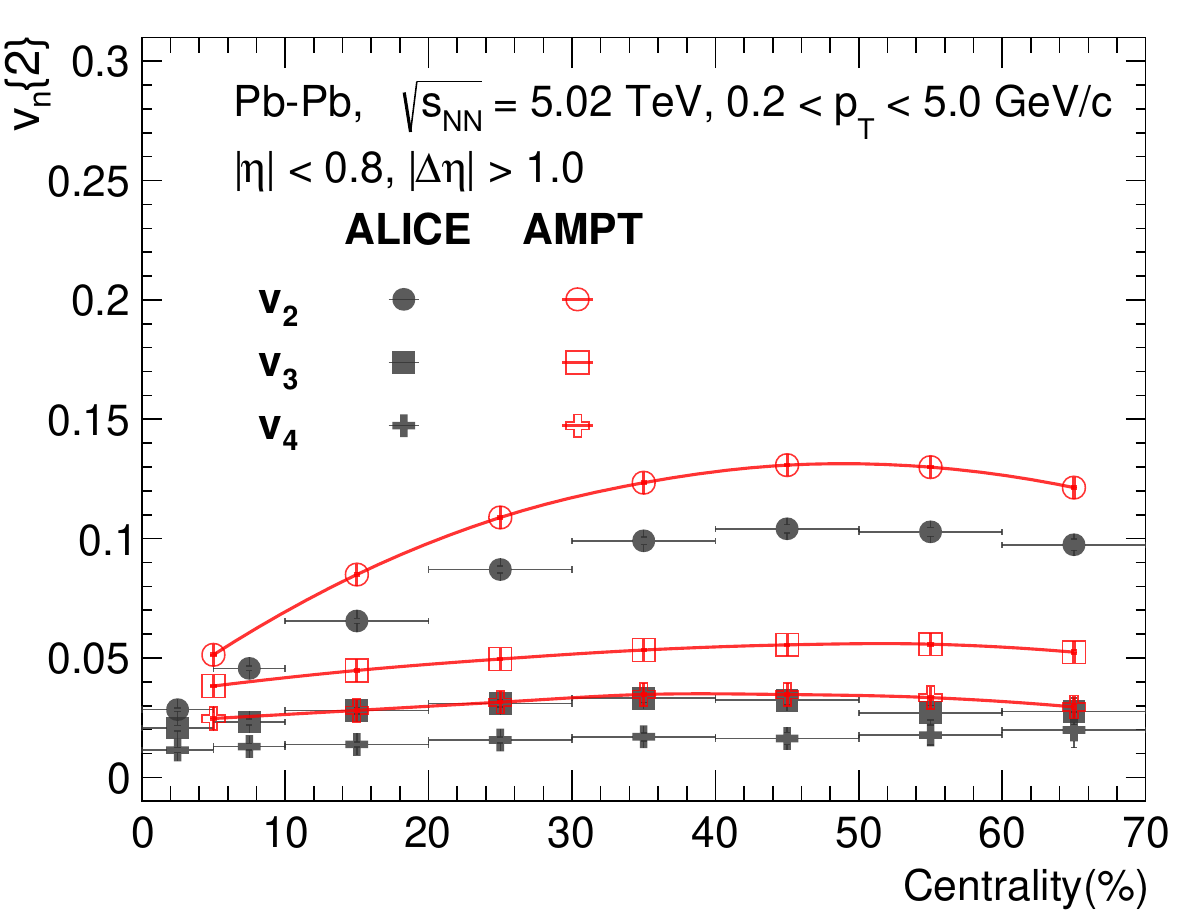}
\caption{Centrality dependence of $v_{n}\{2\}$ of all charged hadrons in Pb--Pb collisions at $\sqrt{s_{\rm NN}}=5.02$ TeV using AMPT compared with corresponding ALICE measurements~\cite{ALICE:2016ccg}.}
\label{fig:ALICEcomp}
\end{figure}

\begin{figure*}[ht!]
\centering
\includegraphics[scale=0.21]{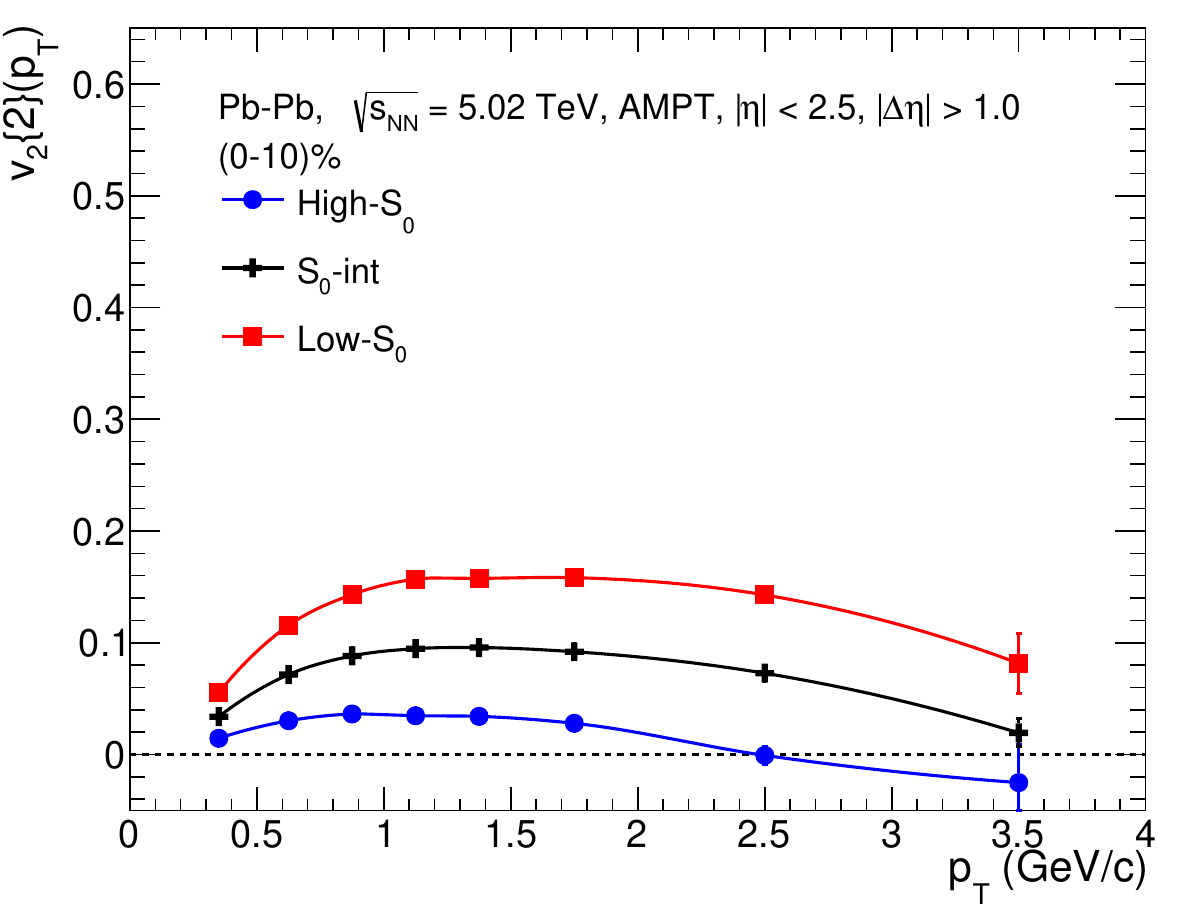}
\includegraphics[scale=0.21]{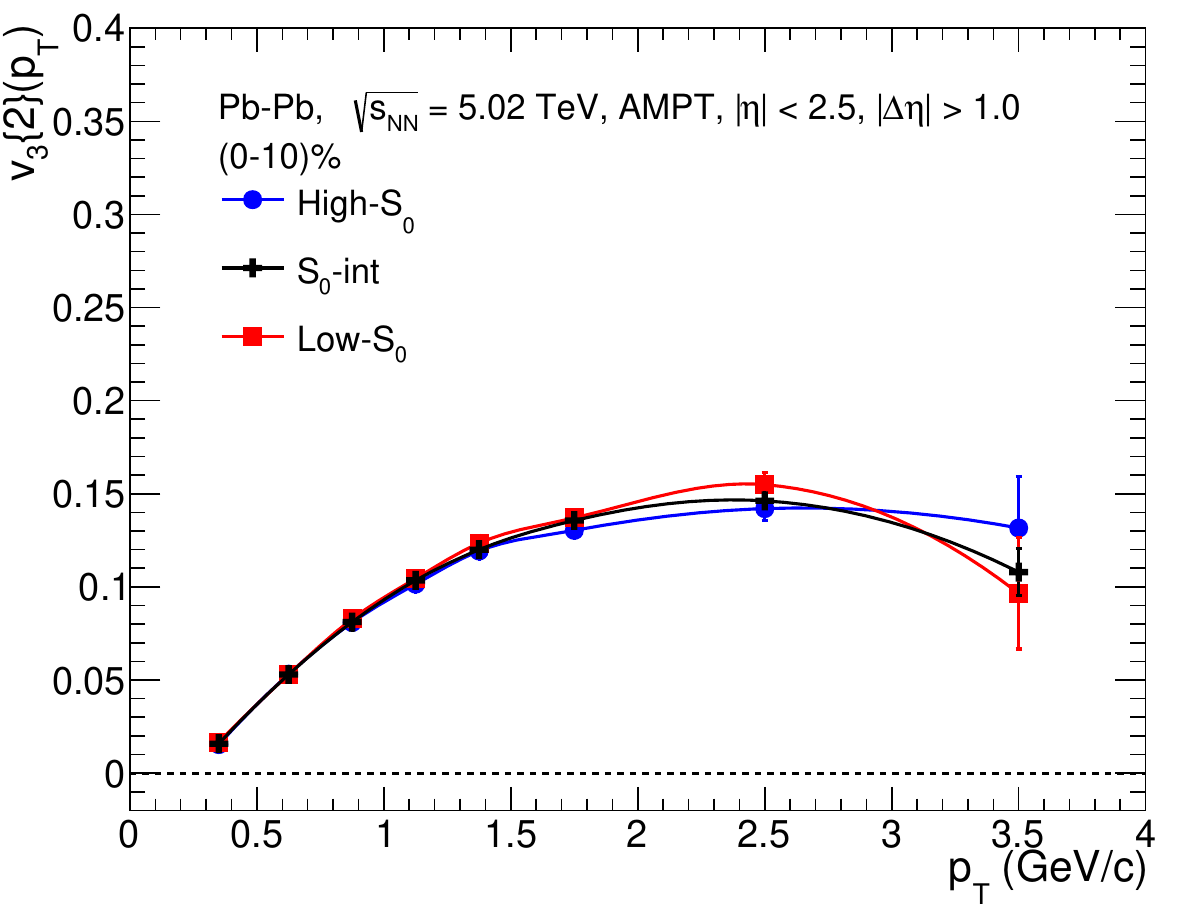}
\includegraphics[scale=0.21]{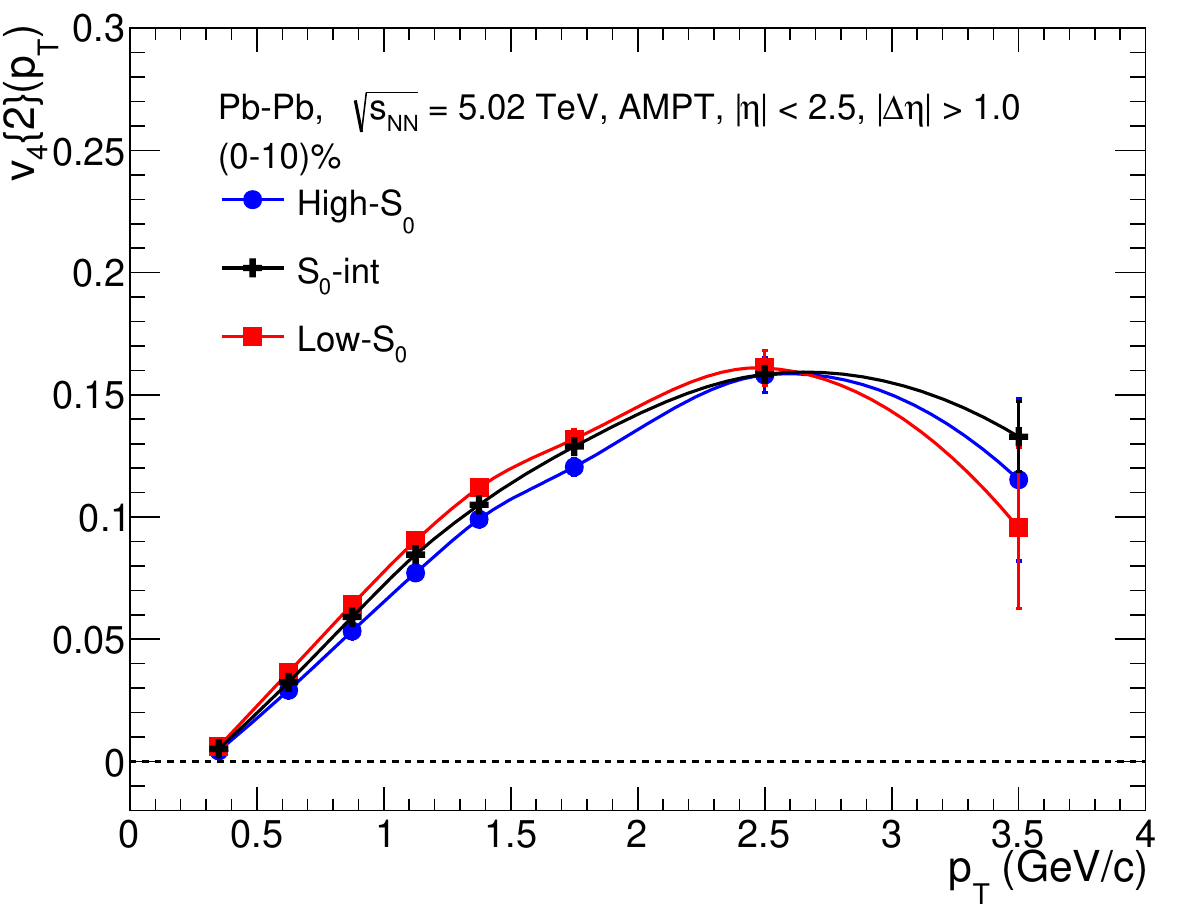}
\includegraphics[scale=0.21]{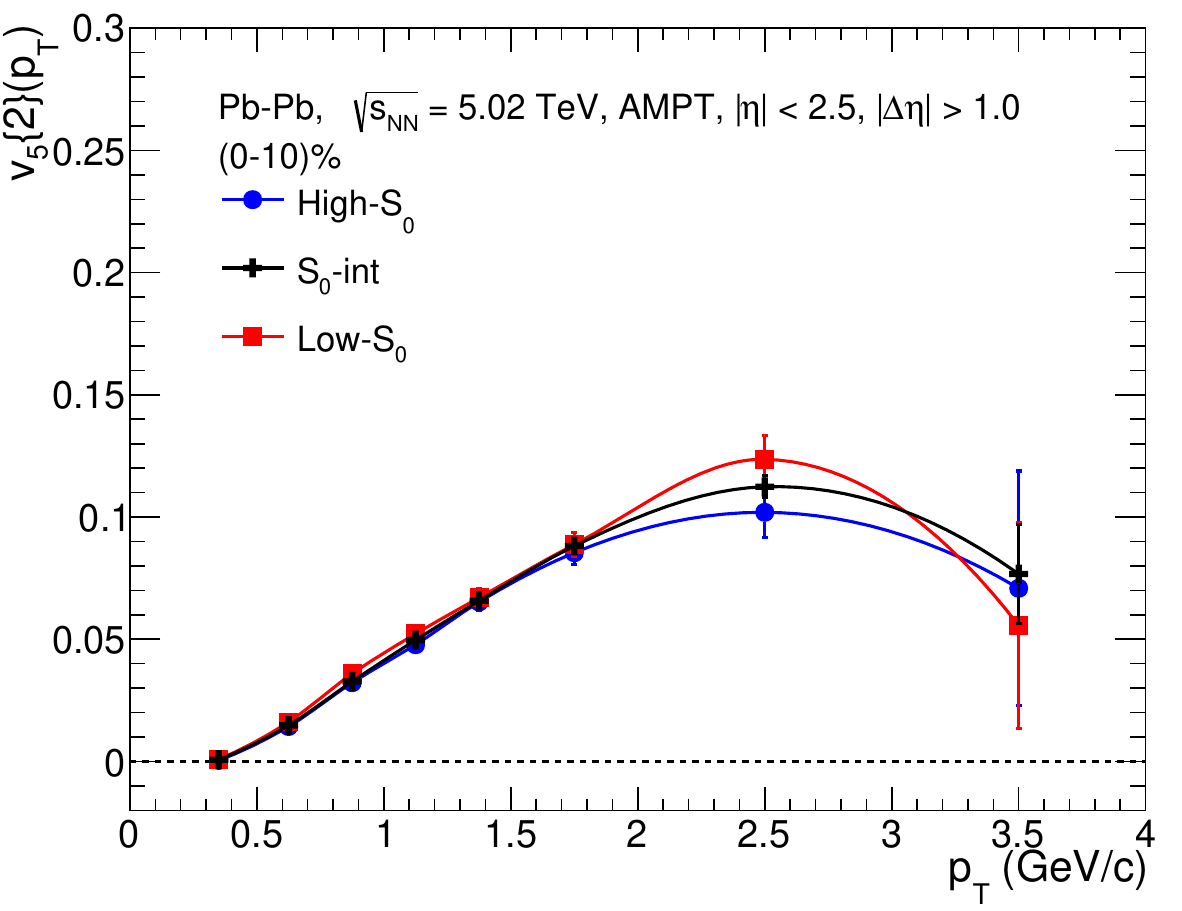}
\includegraphics[scale=0.21]{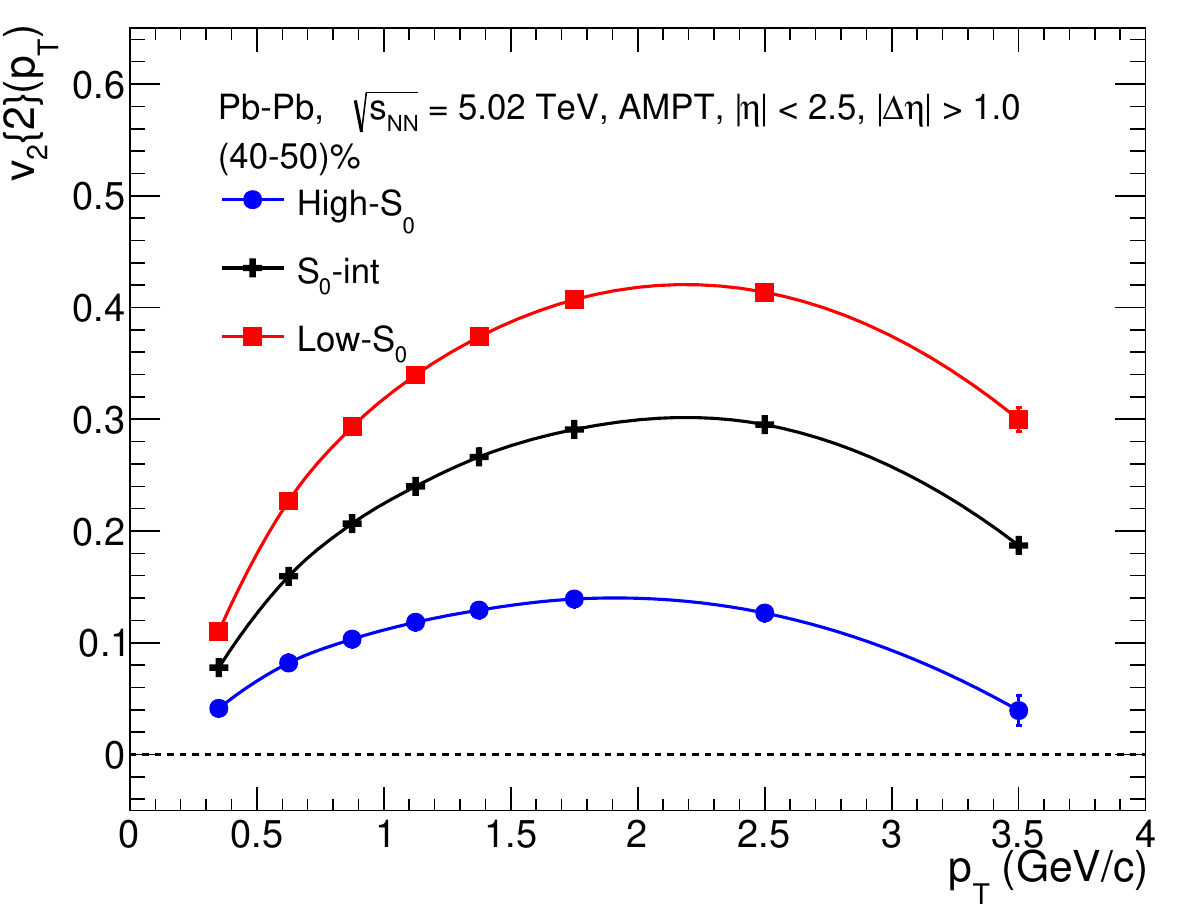}
\includegraphics[scale=0.21]{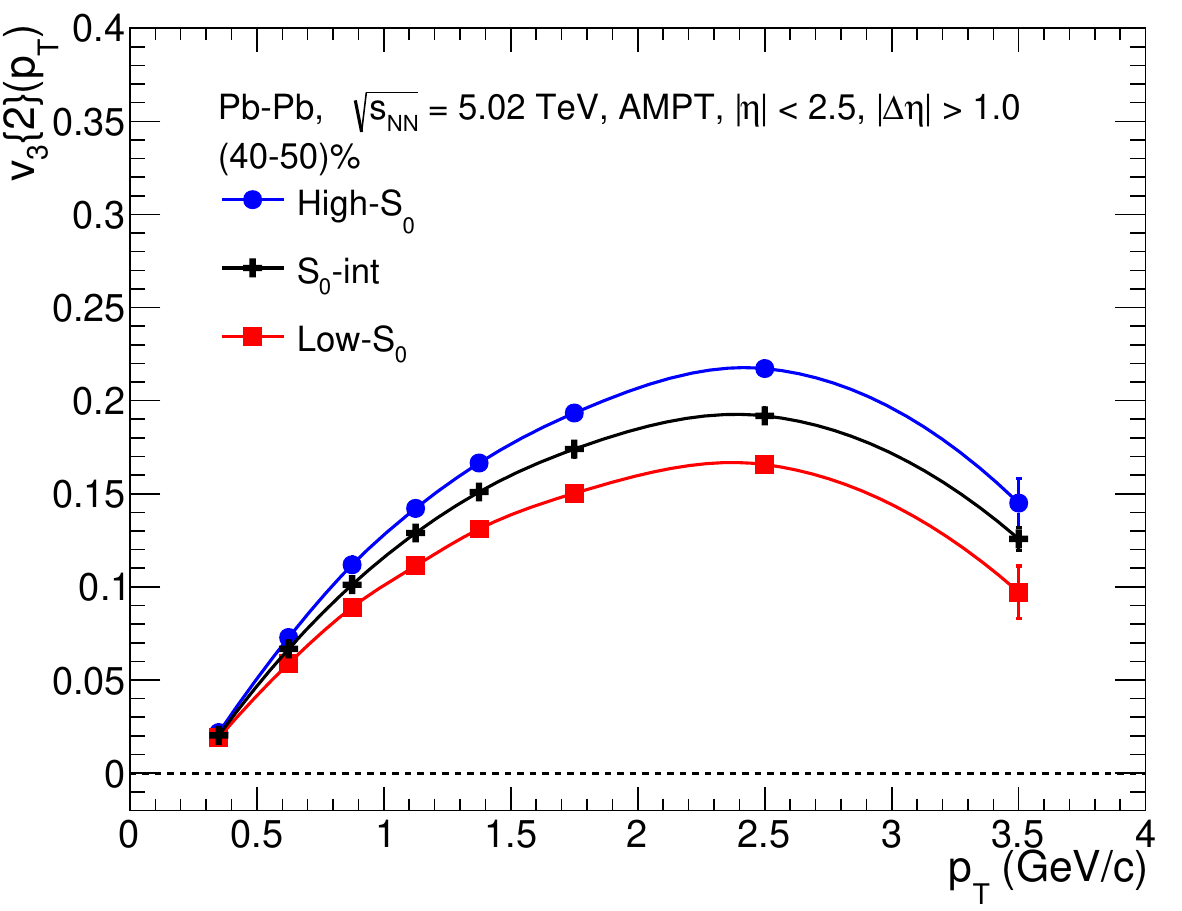}
\includegraphics[scale=0.21]{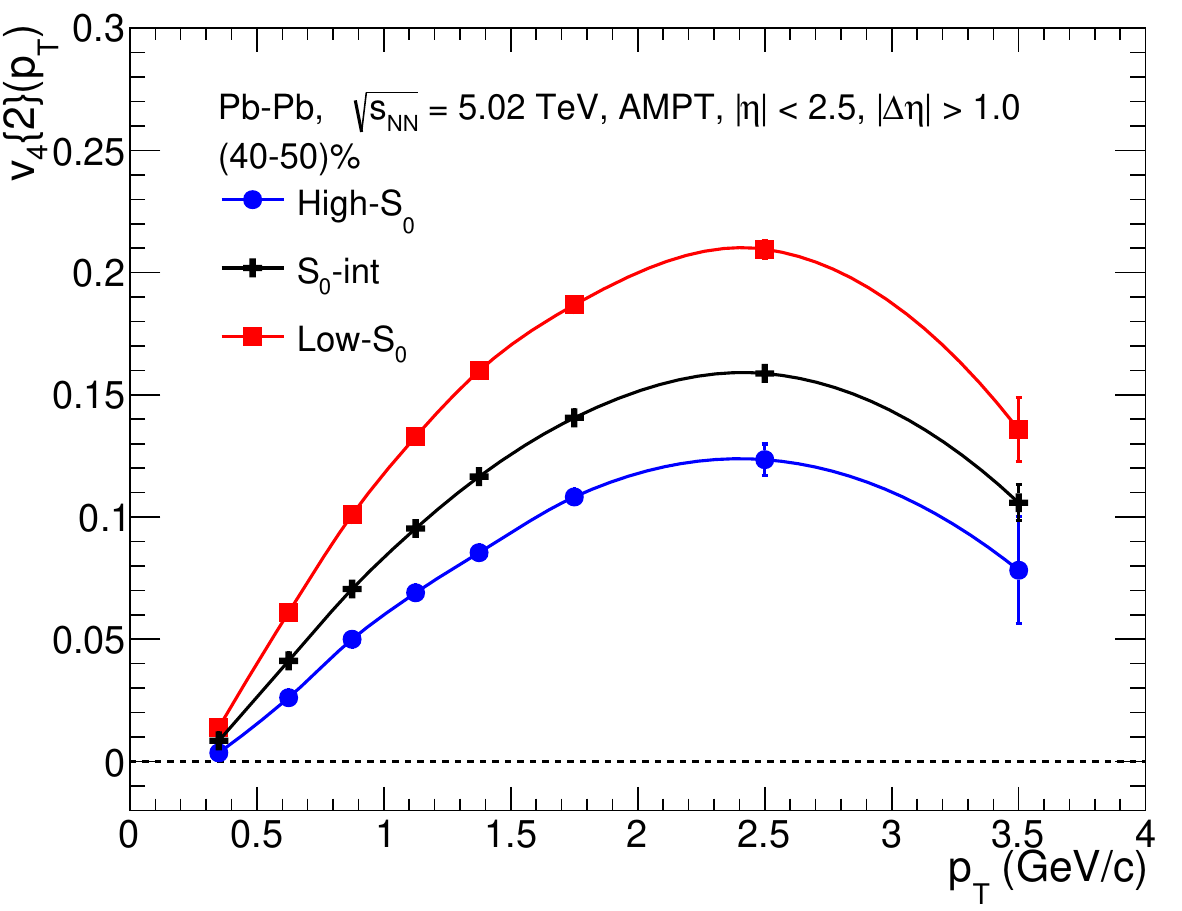}
\includegraphics[scale=0.21]{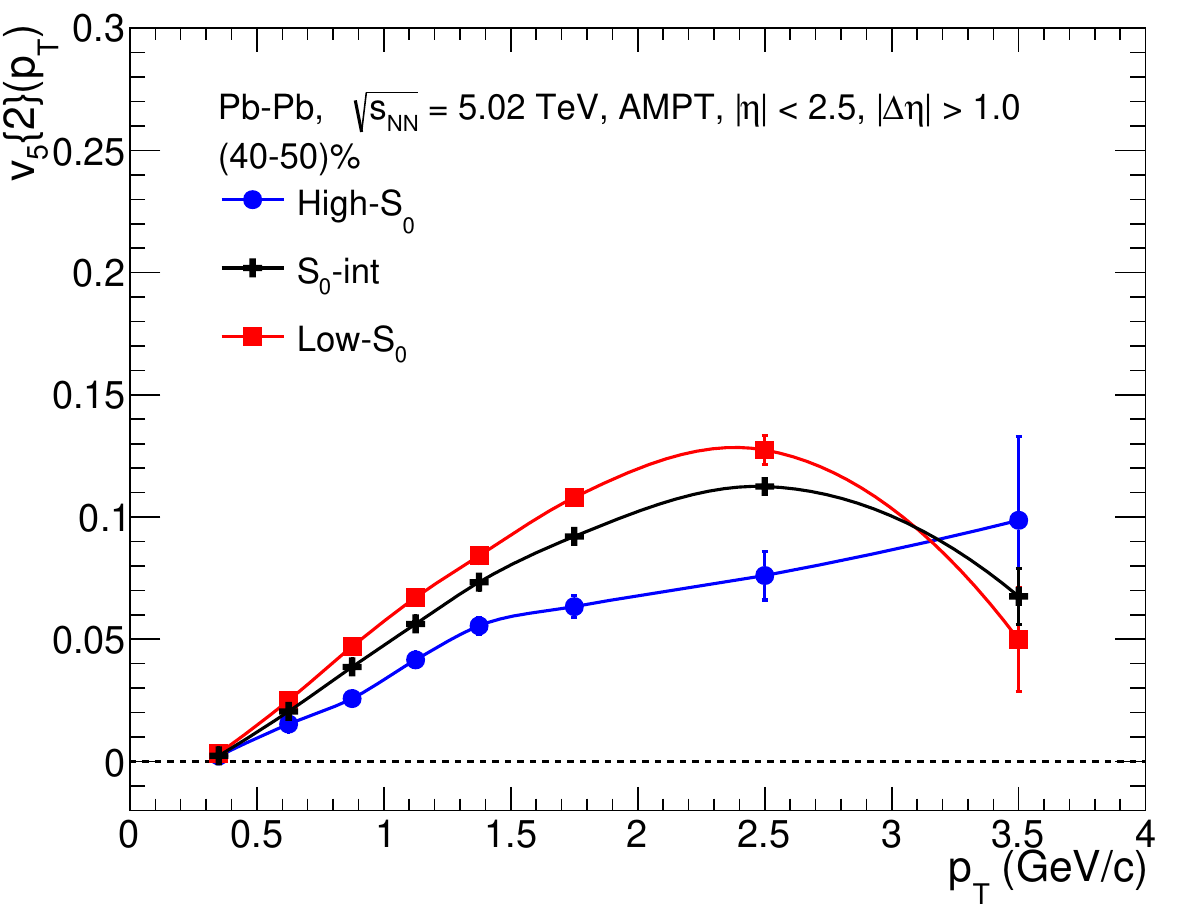}
\caption{$p_{\rm T}$-differential $v_{2}\{2\}$, $v_{3}\{2\}$, $v_{4}\{2\}$ and $v_{5}\{2\}$ (from left to the right) in (0-10)\% (upper) and (40-50)\% (lower) centrality classes for high-$S_0$, $S_{0}$-integrated and low-$S_{0}$ events in Pb--Pb collisions at $\sqrt{s_{\rm NN}}=5.02$ TeV using AMPT.}
\label{fig:vnvspt}
\end{figure*}
\subsection{Comparison with experiments}
Figure~\ref{fig:ALICEcomp} shows the centrality dependence of $v_{2}\{2\}$, $v_{3}\{2\}$ and $v_{4}\{2\}$ measured for particles with $0.2~<~p_{\rm T}~<~5.0$~GeV/c and $|\eta|<0.8$ using two particle Q-cumulant method having two subevents separated by $|\Delta\eta|>1.0$ in Pb--Pb collisions at $\sqrt{s_{\rm NN}}=5.02$ TeV using AMPT. The results from AMPT are also compared with corresponding measurements from the ALICE experiment~\cite{ALICE:2016ccg}. One observes that the calculations of anisotropic flow coefficients using AMPT simulations over-predict the experimental measurements. However, the centrality dependence of $v_{2}\{2\}$, $v_{3}\{2\}$, and $v_{4}\{2\}$ using AMPT is closer in trend to that in experiments. 

\subsection{$p_{\rm T}$ dependence of $v_n$}
Figure~\ref{fig:vnvspt} shows the transverse momentum dependence of anisotropic flow coefficients, $v_{2}\{2\}$, $v_{3}\{2\}$, $v_{4}\{2\}$, and $v_{5}\{2\}$ (from left to right) in (0-10)\% (upper) and (40-50)\% (lower) centrality classes for high-$S_{0}$, $S_{0}$-integrated and low-$S_{0}$ events in Pb--Pb collisions at $\sqrt{s_{\rm NN}}~=~5.02$~TeV using AMPT. Here, the anisotropic flow coefficients increase from low-$p_{\rm T}$ region to intermediate $p_{\rm T}$ regions, followed by a decrease towards the high-$p_{\rm T}$ regions. This picture of the evolution of $v_{n}\{2\}$ with increasing $p_{\rm T}$ from AMPT is consistent with experimental measurements~\cite{ALICE:2016ccg}. For the (40-50)\% centrality class, where the anisotropic flow coefficients are large, as shown in Fig.~\ref{fig:vnvscent}, one finds a similar transverse spherocity dependence as observed in Fig.~\ref{fig:vnvscent}. Here, $v_{2}\{2\}$, $v_{4}\{2\}$, and $v_{5}\{2\}$ show anti-correlation with event selection based on transverse spherocity, while $v_{3}\{2\}$ shows a positive correlation. In comparison, for (0-10)\% centrality class, except for $v_{2}\{2\}$, no significant transverse spherocity dependence on $v_{n}\{2\}$ versus $p_{\rm T}$ is observed. This is interesting due to the fact that $v_{2}\{2\}$ is driven mostly by the initial geometry of the collision overlap region, while the higher order flow harmonics are related to the density fluctuations. In most central collisions, the observed anisotropic flow coefficients are the consequences of the initial eccentricity and density fluctuations of the participating nucleons, where the effect of transverse spherocity is negligible~\cite{Prasad:2022zbr}. However, due to strong anti-correlation between $v_{2}\{2\}$ and $S_{0}$, one finds the observed effects of $S_{0}$ selection on $v_{2}\{2\}(p_{\rm T})$ which is absent in the higher-order flow harmonics for the central heavy-ion collisions.

\end{document}